\documentclass[article]{aa}
\usepackage[utf8]{inputenc}
\usepackage[LGR,T1]{fontenc}
\usepackage{geometry}
\usepackage{xcolor}
\usepackage{subcaption}
\usepackage{float}
\floatstyle{plaintop}
\restylefloat{table}

\usepackage[left,modulo]{lineno}
\usepackage{setspace}

\usepackage{textcomp}
\usepackage{amstext}
\usepackage{graphicx}
\usepackage{upgreek}
\usepackage{txfonts}
\usepackage{amsmath}
\usepackage[]{hyperref}

\usepackage{lipsum, babel}

\hypersetup{colorlinks=true,citecolor=blue}

\definecolor{crimson}{rgb}{0.75, 0.0, 0.2}

\def\Tmap{0.55}

\usepackage[colorinlistoftodos]{todonotes}

\newcommand{\fig}[1]{Fig.\,\ref{#1}}

\def\rj{$R_\mathrm{J}$}

\def\rs{$R_\mathrm{S}$}

\def\hho{H$_2$O}
\def\hh{H$_{2}$}

\def\COratio{$[\mathrm{CO}]/[\mathrm{H}_2\mathrm{O}]$ }
\def\VOratio{$[\mathrm{CO}]/[\mathrm{VO}]$ }
\def\TiOratio{$[\mathrm{CO}]/[\mathrm{TiO}]$ }
\def\Heratio{$\frac{He}{H_2}$ }

\def\wasp{WASP-121\,b}

\def\redchid{\tilde{\chi}^2}

\newcommand{\taurex}{TauREx\xspace}
\newcommand{\pytmo}{Pytmosph3R\xspace}

\title{Towards multi-dimensional analysis of transmission spectroscopy.
Part II: Day-night induced biases in retrievals from hot to ultra-hot Jupiters}

\titlerunning{Day-night induced biases in retrievals from hot to ultra-hot Jupiters}

\author{William Pluriel\inst{1}
     \and
           J\'{e}r\'{e}my Leconte\inst{1}
        \and 
           Vivien Parmentier\inst{2}
        \and
           Tiziano Zingales\inst{1,3}
        \and
           Aur\'{e}lien Falco\inst{1}
        \and
           Franck Selsis \inst{1}
        \and
           Pascal Bord\'{e} \inst{1}
}
           
\institute{Laboratoire d'astrophysique de Bordeaux, Univ. Bordeaux, CNRS, B18N, all\'{e}e Geoffroy Saint-Hilaire, 33615 Pessac, France
\and
Department of Physics, Oxford University, OX1 2JD, United Kingdom (vivien.parmentier@physics.ox.ac.uk)
\and
Universit\`a di Padova, Dipartimento di Astronomia, vicolo dell’Osservatorio 3, 35122 Padova, Italy
}

\abstract{Hot Jupiters are very good targets for transmission spectroscopy analysis. Their atmospheres have a large scale height implying a high signal to noise ratio. As these planets orbit close to their stars, they often present strong thermal and chemical hetereogeneities between the day and the night side of their atmosphere. For the hottest ones, the thermal dissociation of several species occurs in their atmospheres which leads to a stronger chemical dichotomy between the two hemispheres. It has already been shown that the current retrieval algorithms, which are using 1D forward models, find biased molecular abundances in ultra hot Jupiters. Here, we quantify the effective temperature domain over which these biases are present. We use a set of 12 simulations of typical Hot Jupiters from $T_\mathrm{eq}$\,=\,1000\,K to $T_\mathrm{eq}$\,=\,2100\,K performed with the Substellar and Planetary Atmospheric Radiation and Circulation global climate model and generate transmission spectra that fully account for the 3D structure of the atmosphere with \pytmo. These spectra are then analyzed using the 1D \taurex retrieval code.  We find that for JWST-like data, accounting for non-isothermal vertical temperature profiles is required over the whole temperature range. We further find that 1D retrieval codes start to estimate wrong parameter values for planets with equilibrium temperatures greater than 1400\,K if there are absorbers in the visible (such as TiO and VO for instance) able to create a hot stratosphere. The high temperatures at low pressures indeed entail a thermal dissociation of species which creates a strong chemical day-night dichotomy.
As a by-product, we demonstrate that when using synthetic observations to assess the detectability of a given feature or process using a Bayesian framework (e.g., by comparing the Bayesian evidence of retrievals with different model assumptions), it is valid to use non-randomized input data, as long as the anticipated observational uncertainties are correctly taken into account.
}

\begin{document}
\maketitle
%\linenumbers
%\modulolinenumbers[1]

\section{Introduction}

Transmission spectroscopy makes it possible to detect molecules in exoplanetary atmospheres, measure molecular abundances or set upper limits on them. In this regard, several studies pointed out that the 3D nature of real atmospheres, both for thermal and compositional effects, often needs to be taken into account in the modeling for the inference process to yield accurate estimates \citep{caldas2019,Changeat_2019,MacDonald2020,Lacy2020}. Because the atmospheric region probed by transmission spectroscopy is not as thin as often assumed, measured spectra may be affected by variations through and along the terminator (day-to-night and pole-to-equator gradients).

For Ultra Hot Jupiters (UHJs), \citet{Pluriel2020} demonstrated that \COratio estimated with 1D spherically-symmetric models can be off by several orders of magnitude because of strong day-to-night heterogeneities: \hho{} is thermally dissociated in the hot day side, while it is not in the cool night side. Contrarily, the CO abundance remains constant everywhere since temperature gets nowhere high enough for CO to be thermally dissociated. For these planets, transmission spectra will be the combined result of hot regions in the CO bands and cold regions in the \hho{} bands. To capture this complexity, it is necessary to resort to models with more that one dimension. This is especially important in the context of the James Webb Space Telescope (JWST) and Ariel (ESA) that aims at measuring atmospheric elemental abundances with high accuracies \citep{Tinetti2021redbook}.

Hot Jupiters (HJs) and UHJs have been studied with Global Climate Models (GCMs) \citep{Showman2002,Showman2008,Menou2009,Wordsworth2011,Heng2011,Charnay2015,Kataria2016,Drummond2016,TK19}. GCMs are designed to describe planetary atmospheres with their full three-dimensional structures and dynamics, hence leading to deeper insights in their rich physics and chemistry \citep{Showman2008,Leconte2013,Guerlet2014,Venot2014,parmentier2018}. 
We know from GCMs that the cooler the planet the weaker its atmospheric day-night dichotomy with corresponding 3D biases. Somewhere between UHJs and cold planets lies a boundary. It is the aim of this work to assess the equilibrium temperature regions where 1D spherically symmetric models can be safely used -- and where they should not be used -- for the purpose of estimating physical and chemical parameters. To investigate this limit, we have designed and carried out a numerical experiment to identify the origin of the biases and to quantify them.

In the following, we produce synthetic data with a realistic 3D forward model chaining the Substellar and Planetary Atmospheric Radiation and Circulation global climate model (SPARC/MIT) \citep{Showman2008, Showman2013} with our 3D transmission spectrum computation module, \pytmo \citep{caldas2019,falco2021taurex}. Then, we use \taurex \citep{Waldmann2015} to solve the inverse problem with a 1D forward model and estimate the parameters of interest. We will refer to this last part as the retrieval process or retrieval for short, since we investigate if correct parameter values are retrieved from the synthetic data. In this last part, we adopt the point of view of an observer.

In Sec.~\ref{num exp}, we explain how we built our numerical models and the parametrization we used. In Sec.~\ref{noise analysis}, we discuss the use of non-randomized input data and its validity to compare various retrieval hypothesis. In Sec.~\ref{input}, we describe the GCM outputs and the transmission spectra generated with \pytmo, which is used as an input for the retrieval analysis with \taurex (see Sec.~\ref{retrieval}). Finally, we discuss in Sec.~\ref{discussion} how 3D effects (vertical, across and along the limb) alter transmission spectra and vary as a function of equilibrium temperature and the nature of optical absorbers.

\section{Numerical experiment}
\label{num exp}

\subsection{Global Climate Model}
\label{GCMs}

\begin{figure}
\centering
\includegraphics[scale=0.55]{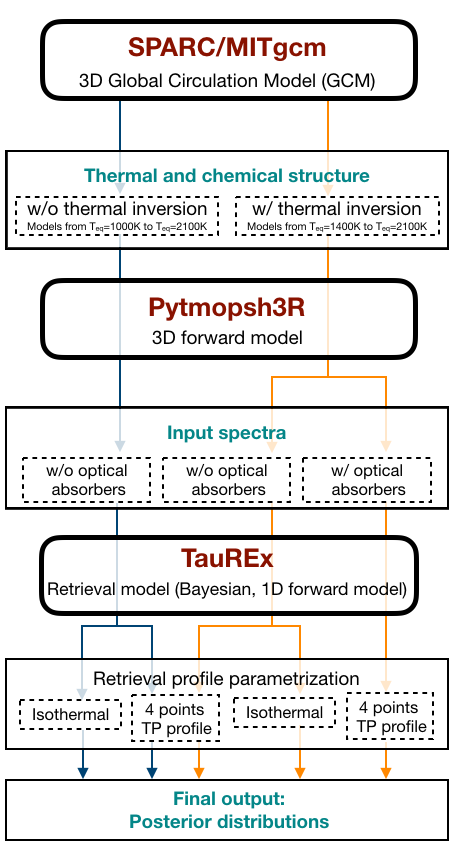}
\caption{Flowchart of the numerical experiment. The 3 models are shown in red, the output are indicated in green-blue and the parameters are presented in black. There are 2 main sets of simulations from the Global Climate Model SPARC/MIT, with and without thermal inversion, respectively the orange and the blue arrows. The final output consists of 5 sets of posterior distributions.}
\label{flowchart}
\end{figure}

To implement our numerical experiment (Fig.~\ref{flowchart}), we followed the methodology described by \citet{Pluriel2020}. In a first step, we ran SPARC/MIT simulations of HJs with equilibrium temperatures between 1000 and 2100 K as in \citet{Parmentier2021}. We decided to truncate at 2100 K because we already shown in \citet{Pluriel2020} the limitation of 1D retrieval model on a 2350 K equilibrium temperature planet (WASP-121\,b). As we already
demonstrated the limitation for hottest planet, we interested here to colder planets. In each simulation, the pressure ranged from $2.0\,10^7$ to 0.2~Pa over 53 levels. We used a horizontal resolution of C32, meaning that each of the six "cube faces" has a resolution of 32$\times$32 finite volume elements \citep{Showman2009}, equivalent to an approximated resolution of 128 cells in longitude and 64 in latitude. Radiative transfer was handled with a two-stream radiation scheme \citep{Marley1999}. Opacities were treated using eight correlated-k coefficients \citep{Goody1989} within each of 11 wavelength bins \citep{Kataria2013}, assuming chemical equilibrium. To ease comparisons, we used the following planetary and stellar parameters common to all simulations:
\begin{itemize}
\item[•] $T_*$ = $T_S$ = 5778 K,
\item[•] $R_*$ = \rs = 6.957$\times10^8$ m,
\item[•] $R_p$ = \rj = 7.1492$\times10^7$ m,
\item[•] Surface gravity: $g$ = 10 m.s$^{-2}$,
\item[•] \Heratio = 0.25774,
\item[•] Solar metallicity,
\item[•] Chemical equilibrium.
\end{itemize}
Thus, from one simulation to the next, only the chemical composition and the equilibrium temperature vary. This approach makes it possible to isolate different effects and facilitates interpretation. We thus do the hypothesis that the parameters chosen fit for different planet type from hot to ultra hot Jupiter, which is consistent with the current observations of this range of exoplanets. Also, we assume a G-type star emitting like a black-body. A different spectral type of the host star would affect the transmission spectra even if the G-star black-body hypothesis is a fair assumption. For instance, we do not observe hot and ultra hot Jupiters around M dwarf which have indeed more complex spectral type with strong lines in the UV.

The strong atmospheric dichotomy in UHJs is not only due to extreme irradiation, but also to absorbers in the near UV and optical domains, such as TiO and VO \citep{Fortney_2008,Parmentier2015,parmentier2018}, causing both a strong heating and a stratospheric thermal inversion. Although These molecules can get depleted by cold traps \citep{Spiegel_2009_can_tio} in cool planets, it will not happen in the day sides of UHJs. Therefore, thermal inversion is very likely in HJ atmospheres. It is however possible that TiO/VO rainout. Thus, the upper atmosphere will be depleted in TiO/VO according to the equilibrium \citep{Parmentier_2016}. To consider this possibility, we chose to simulate planets with and without TiO and VO for this study. Thus, we ran two sets of simulations: 12 \textit{No thermal inversion} simulations without TiO and VO, and with equilibrium temperatures ranging from 1000 K to 2100 K in steps of 100 K, and 8 \textit{Thermal inversion} simulations with TiO and VO, and equilibrium temperatures ranging from 1400 K (minimal value ensuring gaseous forms for TiO and VO) to 2100 K in steps of 100 K. Although we did not include these species, we note that metals such as Fe or Mg, and ionized hydrogen, might also contribute significantly to thermal inversion in UHJs as pointed out by \citet{Lothringer2018}.

\subsection{Transmission spectra}
\label{sub:transmission-spectra}

In a second step, for each simulated planet, we take the temperature maps and chemical abundance tables produced by our GCM to generate 3D transmission spectra with the new version of \pytmo \citep{falco2021taurex} and using monochromatic cross sections calculated by ExoMol \citep{Yurchenko2011, Tennyson2012, Barton2013, Yurchenko2014, Barton2014}. The spectra are computed according to the two envisioned atmospheric compositions:
\noindent (i) an atmosphere mainly composed of \hh{} and He containing only \hho{} and CO as active gases. To account for thermal dissociation of \hho{} and \hh{}, we used the following equation from \citet{parmentier2018} :
\begin{equation}
    \frac{1}{A} = \left(\frac{1}{A^{0.5}_{0}}+\frac{1}{A^{0.5}_{d}}\right)^2,
    \label{abundances}
\end{equation}
where $A_0$ is the deep abundance, unaffected by dissociation and $A_d$ is the abundance in the region dominated by dissociation. \\
\noindent (ii) same atmosphere as (i) but with added TiO and VO. Condensation of these species are accounted for.

In this work, we assume that observations are carried out with JWST and we assume that measurement noise is dominated by quantum detection noise following a Poisson statistics with equal mean and variance between $\sim$2 and 10~$\muup$m. We estimate the mean number of collected photo-electrons to be
\begin{equation}
N_\mathrm{ph} = \frac{\pi \tau \Delta t}{hc} \left( \frac{R_\star D}{2d}\right)^2 \int_{\lambda_1}^{\lambda_2} B(\lambda, T_\star)\lambda d \lambda,
\label{photon_noise}
\end{equation}
where $\lambda_1$ and $\lambda_2$ are the limiting wavelengths of the spectral bin, $d$ is the distance of the star (we took here 270 pc as if it was WASP-121), and $R_\star$ and $T_\star$ are respectively the stellar radius and temperature. The parameters $D$, $\tau$, and $\Delta t$ are respectively the telescope diameter, the system throughput, and the integration time, whose values have been fixed for JWST according to \citet{Cowan2015}. From 0.6 to $\sim 2 \: \muup$m, since systematics may prevent us from reaching a 10 ppm precision with JWST, wherever quantum noise was lower than 30 ppm, we assumed a floor noise of 30 ppm throughout the whole spectral domain with a normal distribution \citep{Greene_2016}. The noise depends on the wavelength above 2 microns according to eq~\ref{photon_noise}, it reaches around 100 ppm at 10 microns. We simulated JWST observations from 0.6 to 10~$\muup$m using the low resolution prism mode provided by the Near-Infrared Spectrograph (NIRSpec) and the Mid Infra-Red Instrument (MIRI) \citep{Stevenson_2016}. We are aware that this observational set-up is somewhat ideal but what we seek here is to uncover biases due to atmospheric modeling, not due to instrumental effects. We specify that we used the standard deviation of the quantum noise as the estimate of the uncertainty affecting the spectrum but we did not actually add random noise to the spectra we computed. In Sec.~\ref{noise analysis}, we present a study checking that using non-randomized spectra does not affect our conclusions.
%We specify here that we treat the stellar noise as uncertainties added on the generated transmission spectra, but spectra will not be noised. 
%We take advantage of the fact that, as shown by \citet{Feng2018}, retrievals converge to the nonrandomized posterior distribution, no matter the actual noise on the spectra. We tested that we indeed found the same posterior distribution with noisy and nonrandomized spectra specifying the same uncertainties. Using this nonrandomized configuration allows the \redchi\ to be smaller than 1. Values of \redchi\ below 1 all correspond to an equivalent highest level of agreement between the observation and the model.

\subsection{Parameter retrieval}
\label{sub:parameter-retrieval}

In a third step, we used \taurex for retrieving the atmospheric parameters listed in Table~\ref{tab: priors}, assuming uninformative flat priors with broad ranges, exactly as an observer would do. We ran the retrieval process twice, first with an isothermal profile, then with a 4-point (6-parameter) thermal profile. We did not include clouds, neither in the first, not in the second step of the numerical experiment. It would however be interesting to add clouds in the simulations. As the clouds would affect mostly the short wavelengths, we think that the biases in the CO/H$_2$O ratio we observe would still occur since the CO bands are at longer wavelength. But the presence of clouds could create degeneracies in the retrieval. Clouds are nevertheless part of the retrieval parameters in the \taurex model. This can break up possible degeneracies and we ensure that the model works correctly by not retrieving a cloud layer when we know that none has been implemented. When performing retrievals, we impose a limiting condition to maintain physical scenarios: we set the cloud pressure range to be between the bottom and the top of the atmosphere.
Finally, we compared the parameter inferred values with their actual values.

\begin{table}
\centering
\begin{tabular}{lll} 
\hline\hline
    Parameters  & W/ thermal inversion & W/O thermal inversion \\ \hline\hline
    $[\mathrm{H}_2\mathrm{O}]$  & 10$^{-12}$ - 10$^{-1}$  & 10$^{-12}$ - 10$^{-1}$ \\
    $[\mbox{CO}]$  & 10$^{-12}$ - 10$^{-1}$  & 10$^{-12}$ - 10$^{-1}$ \\
    $[\mbox{TiO}]$  & 10$^{-12}$ - 10$^{-1}$  & - \\
    $[\mbox{VO}]$  & 10$^{-12}$ - 10$^{-1}$  & - \\
    $R_p\, (R_{jup})$ & $\pm 50\%$ & $\pm  50\%$ \\ 
    $P_{clouds}$ [Pa] & 10$^{-2}$ - 10$^{6}$ & 10$^{-2}$ - 10$^{6}$  \\
    \hline
    \multicolumn{3}{c}{Isothermal profile}\\\hline
    $T_p$ [K] & 0.3$\times T_{eq}$ - 2$\times T_{eq}$ & 0.3$\times T_{eq}$ - 2$\times T_{eq}$ \\
    \hline
    \multicolumn{3}{c}{4-points TP profile}\\\hline
    $T^{bot}$ [K] & 0.3$\times T_{eq}$ - 2$\times T_{eq}$ & 0.3$\times T_{eq}$ - 2$\times T_{eq}$ \\
    $T_{1}$ [K] & 0.3$\times T_{eq}$ - 2$\times T_{eq}$ & 0.3$\times T_{eq}$ - 2$\times T_{eq}$ \\
    $P_{1}$ [Pa] & 10$^{6}$ - 10$^{2}$ & 10$^{6}$ - 10$^{2}$ \\
    $T_{2}$ [K] & 0.3$\times T_{eq}$ - 2$\times T_{eq}$ & 0.3$\times T_{eq}$ - 2$\times T_{eq}$ \\
    $P_{2}$ [Pa] & 10$^{6}$ - 10$^{0}$ & 10$^{6}$ - 10$^{0}$ \\
    $T_{top}$ [K] & 0.3$\times T_{eq}$ - 2$\times T_{eq}$ & 0.3$\times T_{eq}$ - 2$\times T_{eq}$ \\
    \hline
    \end{tabular}
    \caption{Retrieval parameters$^a$.}
    \footnotesize{$^a$ List of retrieval parameters with corresponding fitting ranges. Priors are assumed to be flat in log space for abundances, in linear space otherwise. $T_{1}$ and $T_{2}$ are the temperature points corresponding to pressure points $P_{1}$ and $P_{2}$, respectively.}
    \label{tab: priors}
\end{table}

\subsection{On the use of non-randomized spectra}
\label{noise analysis}

As mentioned in Sec.~\ref{sub:transmission-spectra}, we computed non-randomized spectra, i.e., with no random noise added. The reason is that a particular instance of added random noise would randomly affect the results of our retrieval studies in a particular way, whereas we seek to uniquely identify biases of 1D-model based retrievals. An approach could have been to perform a series of retrievals on a series of instances of added random noise. However, as showed by \citet{Feng2018}, this would have been unnecessarily computationally expensive as the posterior probability distributions of all these combined retrievals would have converged to the posterior distributions obtained from the non-randomized spectrum.

In Sec.~\ref{sub:parameter-retrieval}, we take the point of view of an observer, thus
as observers would usually do, we introduce the reduced chi-square, denoted by $\tilde{\chi}^2$, a common statistical goodness-of-fit metric defined as follows:
\begin{equation}
  \tilde{\chi}^2 = \frac{1}{N-p}\sum_{i=1}^{N}
  {\left[ \frac{O_i-C_i}{\sigma_i}
  \right]}^2,
\end{equation}
\label{redchitwo}
where $O$ and $C$ are respectively the observed and calculated spectra, $\sigma$ are the uncertainties, $p$ the number of parameters and $N$ the total number of measurements. Note that in the case of non-randomized spectra with uncertainty $\sigma_i$, but no actually added white Gaussian noise with standard deviation $\sigma_i$, $O_i-C_i$ can very well be much smaller than $\sigma_i$. Hence, $\redchid{}$ will not have a statistical mean value of 1. Therefore, in our experiment, $\redchid{} \ll 1$ is to be expected and should not be seen as the sign of noise fitting.

Furthermore, we want to check that we can use logarithmic Bayes factors to compare different forward models (assumptions) using non-randomized input spectra.
For this purpose, we use \pytmo to generate a transmission spectrum of a simple homogeneous atmosphere with a vertical temperature gradient. This non-randomized spectrum serves as a reference. Then, from this spectrum, we generate 20 noisy spectra by adding to each data point a random value drawn from a normal distribution with a 30-ppm standard deviation. We then perform retrievals with \taurex on each of the 21 spectra using two different assumptions on the Temperature-Pressure (TP) profile: (i) an isothermal atmosphere and (ii) a 4-point TP profile atmosphere, as described in Table~\ref{tab: priors}. For all retrievals, \taurex is provided with the same 1-$\sigma$ uncertainty (in this simple case, 30~ppm), whether the input spectrum has been noisy or not.

\begin{figure}
\centering
\includegraphics[scale=0.65,trim = 0cm 0cm 0cm 0cm, clip]{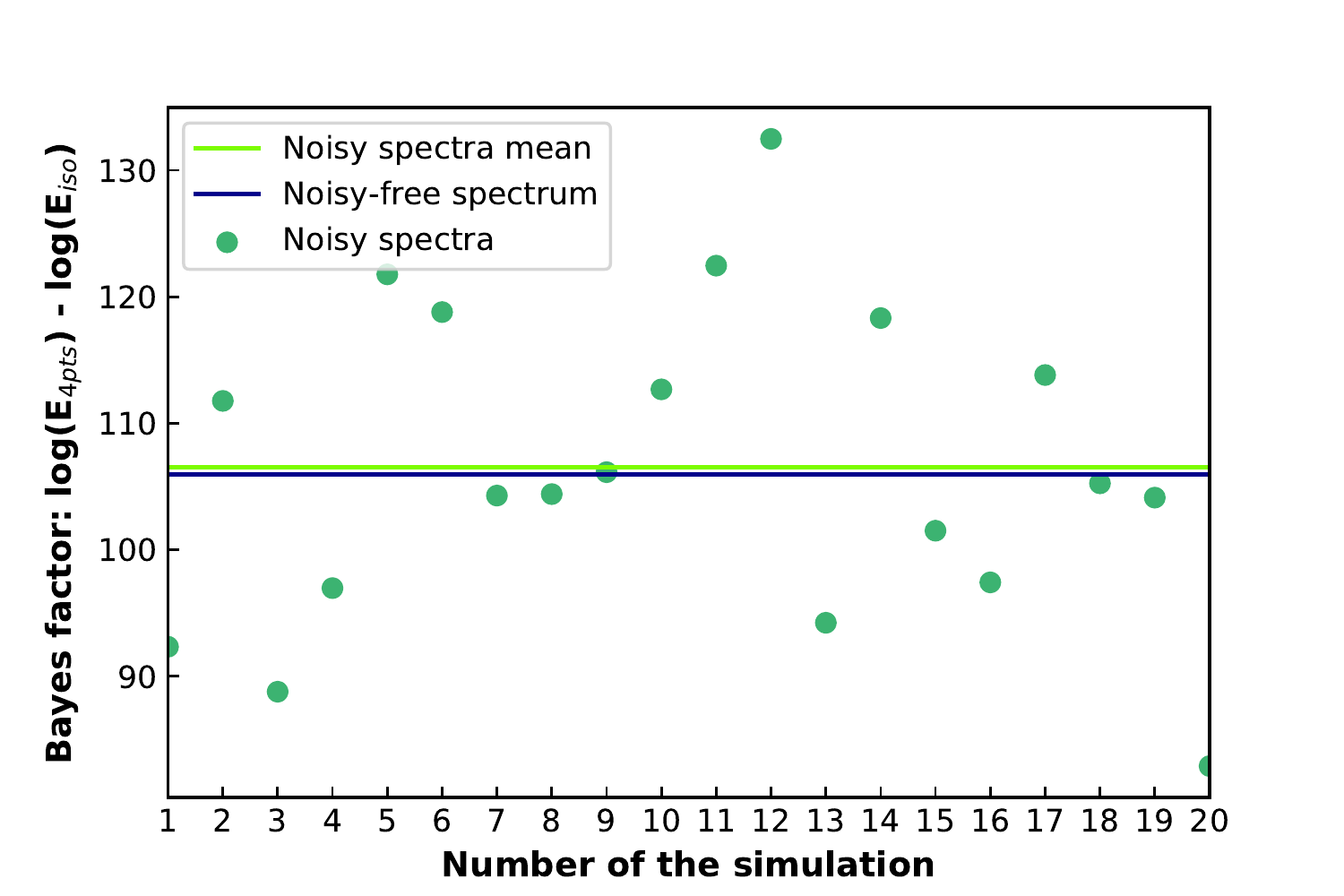}
\caption{Logarithmic Bayes factor for a 4-point TP profile versus an isothermal profile. The figure shows in blue the value obtained with non-randomized spectra, and as green dots the 20 values for normally-distributed noisy spectra (30-ppm standard deviation). The logarithmic Bayes factor averaged on the 20 simulations appears in light green. The non-randomized logarithmic Bayes factor and its noisy-average counterpart are almost identical proving that our study can be based on non-randomized spectra.}
\label{fig: Bayes_factor_analysis}
\end{figure}

We use Bayesian evidences as defined by \citet{Waldmann2015} to compute the logarithmic Bayes factor:
\begin{equation}
  \log B = \log{\frac{E_\mathrm{4pts}}{E_\mathrm{iso}}} = \log{E_\mathrm{4pts}} - \log{E_\mathrm{iso}},
  \label{eq:logbayes_factor}
\end{equation}
where $E_\mathrm{4pts}$ and $E_\mathrm{iso}$ are the evidences of the 4-point TP profile and the isothermal profile, respectively. Fig. \ref{fig: Bayes_factor_analysis} shows the result that strongly favors the 4-point TP profile. Although $\log B$ is widely spread for the randomized spectra, its average value is very close to the non-randomized reference value. We conclude that it is perfectly valid to carry out comparisons of various retrieval hypotheses based on non-randomized data as long as uncertainties are correctly accounted for. In fact, this approach even alleviates potential biases in model selection due to a particular instance of noise. The reason it works is that evidence computation does not rely on the $\redchid$ value of the best fit model, but integrates information over the whole parameter space while always accounting for uncertainty on the data points. Consequently, in the remainder of this article, model selection will always be done by computing logarithmic Bayes factors and never by comparing $\redchid$ values. However, we still keep and indicate $\redchid$ values as potential warnings for unacceptable fits pleading for individual model rejection ($\redchid \gg 1$).
% our logarithmic Bayes factor approach naturally penalizes models with too many free parameters.

While a spread of $\log B$ is to be expected when using randomized data, the spread of about 40 visible in Fig. \ref{fig: Bayes_factor_analysis} appears relatively large compared to the typical value of 5 generally used for model selection. It probably means that selection thresholds for the kinds of retrievals we are dealing with will have to be properly calibrated in the future. 

\section{GCM simulations and transmission spectra}
\label{input}

\subsection{GCM simulations with optical absorbers}

\begin{figure*}
\centering
\includegraphics[scale=0.92,trim = 0cm 14cm 0cm 0cm, clip]{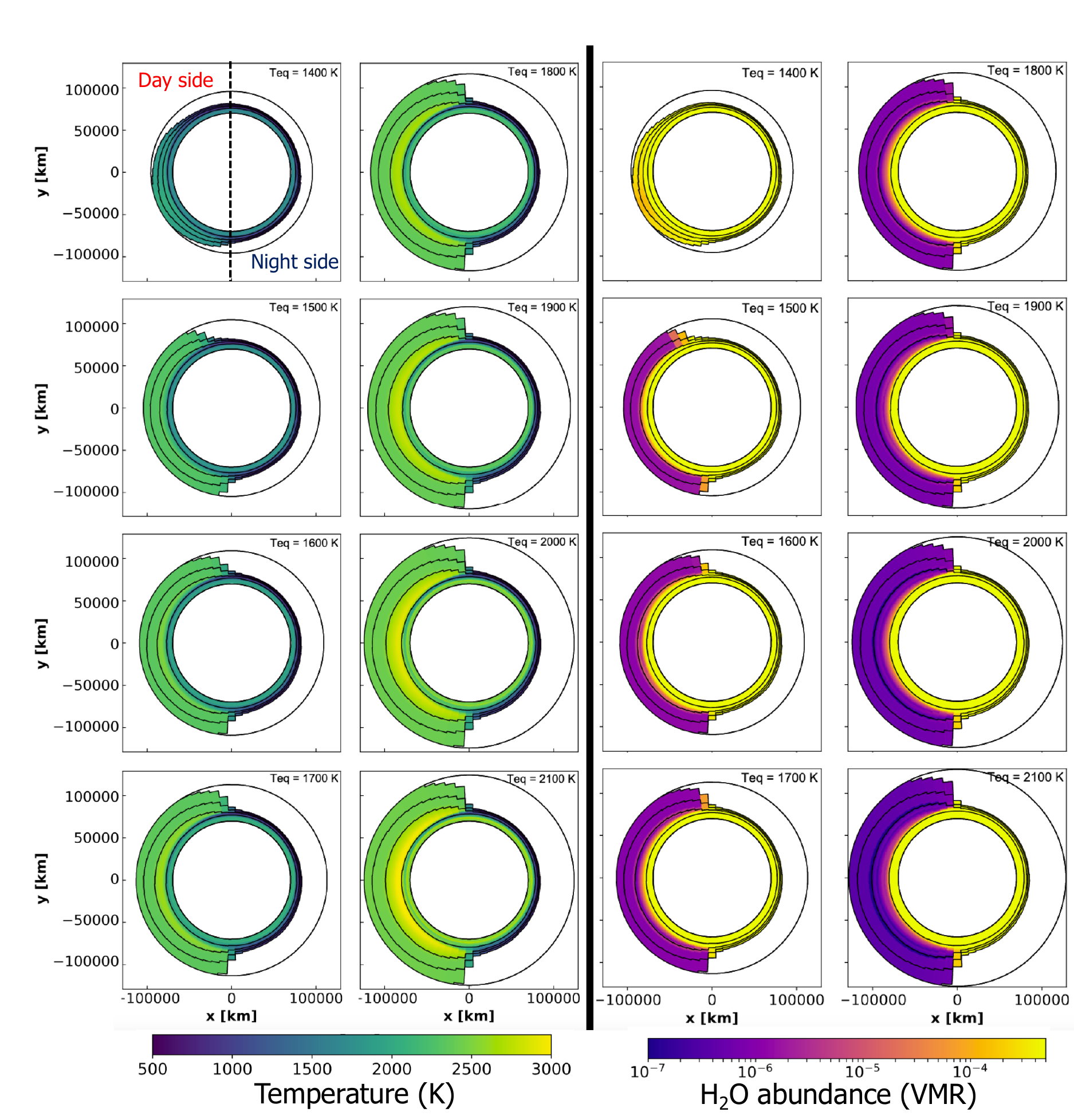}
\caption{Equatorial cut of the temperature and the water abundance %(respectively half left and half right) 
for the 8 GCM simulations with optical absorbers (TiO and VO). The equilibrium temperature of the planet (ranging from 1400\,K to 2100\,K) is specified. From the center outward, the five solid black lines are respectively the $1.434\times10^7$, $10^3$, 1, $10^{-2}$, and $10^{-4}$ Pa pressure levels. The hotter the equilibrium temperature, the larger the day-night thermal and chemical dichotomy. Water is completely dissociated below 10$^3$ Pa in the day side for the hottest simulations.}
\label{fig: maps-TiO}
\end{figure*}

In Fig.~\ref{fig: maps-TiO} we show the temperature and water abundance equatorial maps obtained from GCM simulations with optical absorbers TiO and VO. 
%It appears that the mean temperature of the atmospheres with optical absorbers, increases with the increase of the equilibrium temperatures. \fsnote{What do you mean? When it's hotter, it's hotter? ;) } However, these temperature increases do not affect equally every part of the atmosphere. 
As in the GCM simulation of \wasp{} by \citet{Pluriel2020}, we can distinguish three regions in the atmosphere: i) a quasi-isothermal layer (with only small variations with latitude and longitude) corresponding to the deepest layers of the simulations; above this deep layer, ii) an overall cold region with temperature decreasing with increasing altitude on the night side, and iii) on the day side, a hot stratosphere with the highest temperatures in the simulation.

The temperature of the quasi-isothermal layer increases roughly linearly with the equilibrium temperature of the planet, starting from about 1300\,K ($T_\mathrm{eq}$ = 1400\,K) up to about 2200\,K ($T_\mathrm{eq}$ = 2100\,K). This layer is compressed to higher pressures as $T_\mathrm{eq}$ increases. It extends up to the altitude of the 4\,$10^{3}$ Pa level for $T_\mathrm{eq}$ = 1400\,K, but not above the altitude of the 2\,$10^{4}$ Pa level for $T_\mathrm{eq}$ = 2100\,K.

On the day side, the variation of temperature with latitude and altitude is quite complex. We observe that the location of the hottest region of the atmosphere depends on the equilibrium temperature of the simulation. While it is aligned with the sub-stellar point for the hottest simulations, it is shifted eastward for cooler simulations, up to $23^{\circ}$ for $T_\mathrm{eq}$ = 1400\,K. This well-known shift \citep{Knutson2007,Showman2008} occurs when the energy advection timescale becomes smaller than the radiative timescale. In this case, the hottest point is controlled by both circulation and radiation and is displaced to the east by zonal winds. The more intense and organized the wind dynamics (jets, super-rotation), the greater this shift between the sub-stellar point and the hottest region. However, for the most irradiated planets, the radiative time scale becomes shorter than the dynamics timescale, hence an alignment between the sub-stellar point and the hottest region. Thus, there is an asymmetry between the east and the west of the atmosphere, which extends well beyond the terminator for the colder simulations. The strong dichotomy in day-night temperature of the hottest atmospheres induces wind dynamics that are sufficiently effective to affect the terminator of these planets. Differences in temperatures and therefore in scale heights are clearly visible on the terminator of these atmospheres. The different east and west terminator signatures will thus mix into the global transmission spectra. Nevertheless, these asymmetries disappear very quickly by moving away from the terminator for the hottest planets.

Regarding the overall change in temperature on the day side, it increases sharply as the equilibrium temperature increases, implying a significant increase in the scale height of this hemisphere, much greater than that of the night hemisphere. This stratospheric heating by optical absorbers significantly enhances the day-night asymmetry as the planetary equilibrium temperature raises. The temperature on the night side of each simulation in fact remains very cold overall, and although it increases slightly with the equilibrium temperature, its scale height remains very small compared to that of the day side \citep{Parmentier2021}.

We also show in Fig.~\ref{fig: maps-TiO} volume mixing ratio (VMR) maps of \hho{}. In our hottest simulations, water thermal dissociation is extremely efficient on the day side of the planet, gradually disappearing with decreasing $T_\mathrm{eq}$ and becoming finally negligible at $T_\mathrm{eq} = 1400$~K. As the considered equilibrium temperature decreases, we notice that the dissociation of \hho{}, which depends on both temperature and pressure, affects regions of lower pressures.
%showing that th even though, in these regions it occurs less efficiently. Indeed, since thermal dissociation depends on both, temperature and pressure, this behavior is expected. 
Concerning the abundance of \hho{} at the terminator, it remains close to solar abundance for all simulations. Only the hottest simulations reveal a slight decrease in the abundance of \hho{} for the regions with the lowest pressures, even though it remains above $10^{-4}$ in VMR.

\begin{figure}
\centering
\includegraphics[scale=0.28,trim = 0.3cm 0.2cm 1cm 0.8cm, clip]{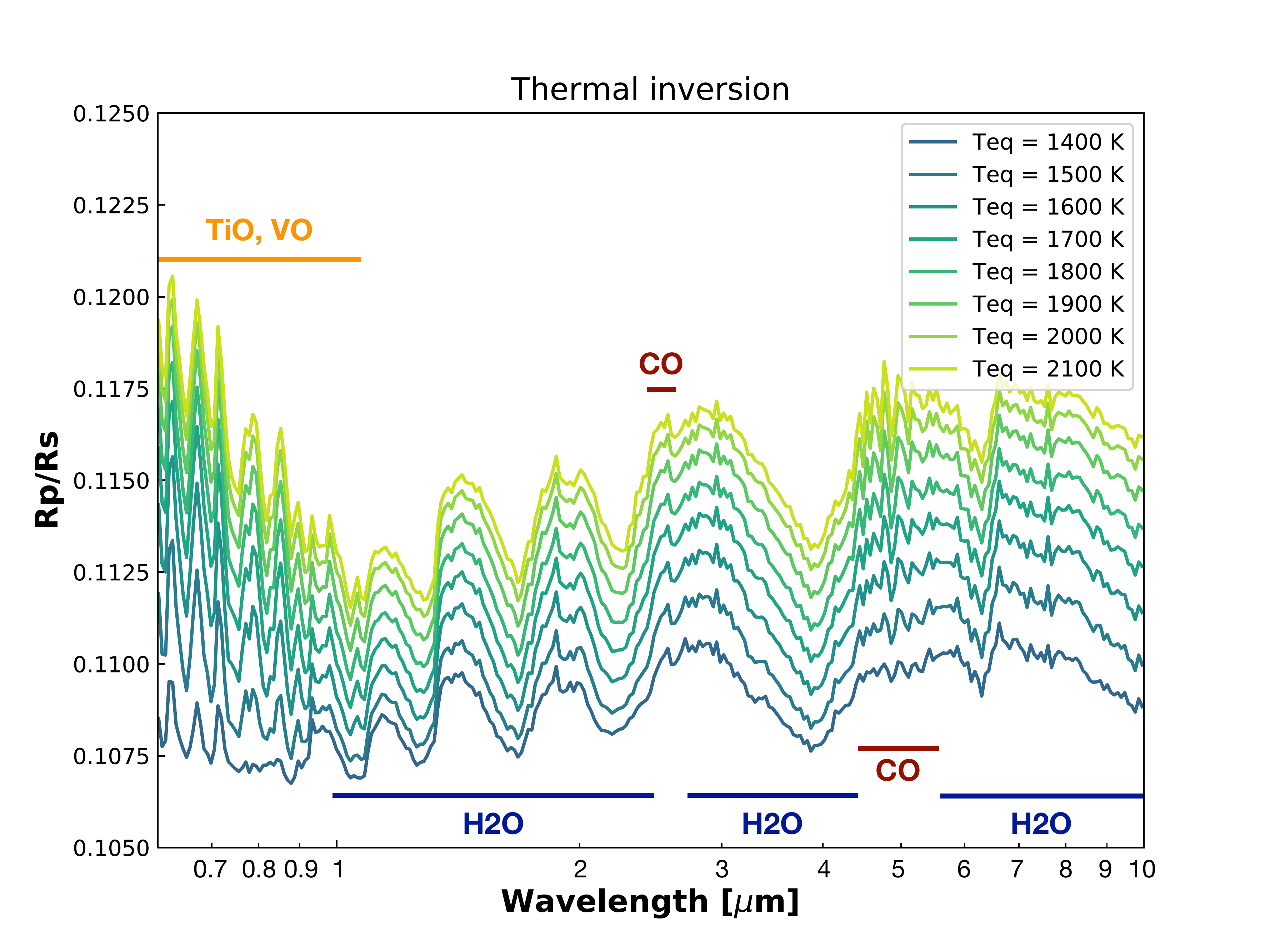}
\caption{Transmission spectra calculated with Pytmosph3R from the 8 GCM simulations with optical absorbers (VO and TiO). We added the main spectral bands of CO, TiO/VO and H$_2$O. CO bands become deeper when temperature increases while the TiO/VO bands become shallower when temperature decreases due to condensation of these species.}
\label{fig: Spectra-GCM}
\end{figure}

Fig.~\ref{fig: Spectra-GCM} shows the spectra generated with \pytmo from our \textit{Thermal inversion} simulations. The transmission spectra are all shifted to larger $R_p$/$R_\star$ ratios for increasingly hotter planets. This vertical shift indicates that higher altitudes are probed, implying more extended atmospheres with larger scale heights. The scale height rises because of temperature increase and molecular weight decrease resulting from thermal dissociation of H$_2$. If water were not thermally dissociated, $R_p$/$R_\star$ would be larger and would hide most of the CO features. However, as \hho{} dissociates in the dayside, the water features come from the nightside while the CO features come from the dayside of the atmosphere, thus they appear more clearly in the spectra. We also see that adding TiO and VO in the atmosphere adds features in the optical domain, hiding the Rayleigh slope and several water bands in the near IR. 
The amplitude of the H$_2$O absorption bands increases with equilibrium temperature. As water absorption increases with temperature \citep{Yurchenko2011}, it indicates that higher temperatures are probed. Moreover, the CO absorption bands (at 2.3 and $4.5\:\muup$m), which remain quite weak for the coldest simulations, appear more clearly for the hottest simulations. Indeed, since CO remains present everywhere in the atmosphere without being dissociated, the absorption bands are dominated by hot day-side features, and this trend increases with the equilibrium temperature as the day-night gradient. This means that when we simulate hotter planets, the greater the equilibrium temperature the greater the difference between the temperature probed by the absorption bands of H$_2$O and those of CO. We also see in Fig. \ref{fig: Spectra-GCM} that the condensation of TiO and VO occurs in the coldest simulation implying a decrease of the features in the optical part of the spectra. However, for the hottest simulation at $T_\mathrm{eq}$ = 2100\,K, the spectra are unaffected by condensation.

\subsection{GCM simulations without optical absorbers}

\begin{figure*}[h!]
\centering
\includegraphics[scale=\Tmap,trim = 5.55cm 6.1cm 7.6cm 4cm, clip]{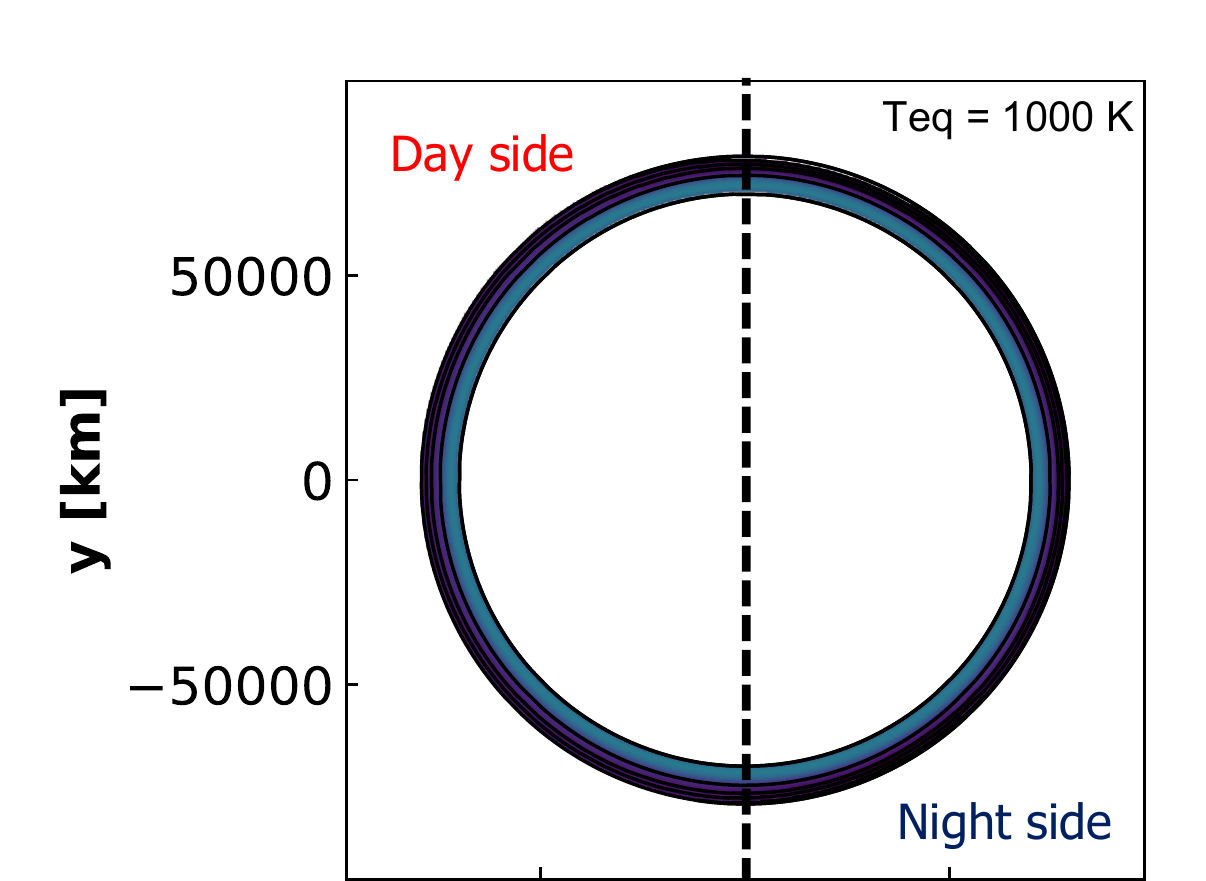}
\includegraphics[scale=\Tmap,trim = 3.9cm 1.65cm 2.5cm 0cm, clip]{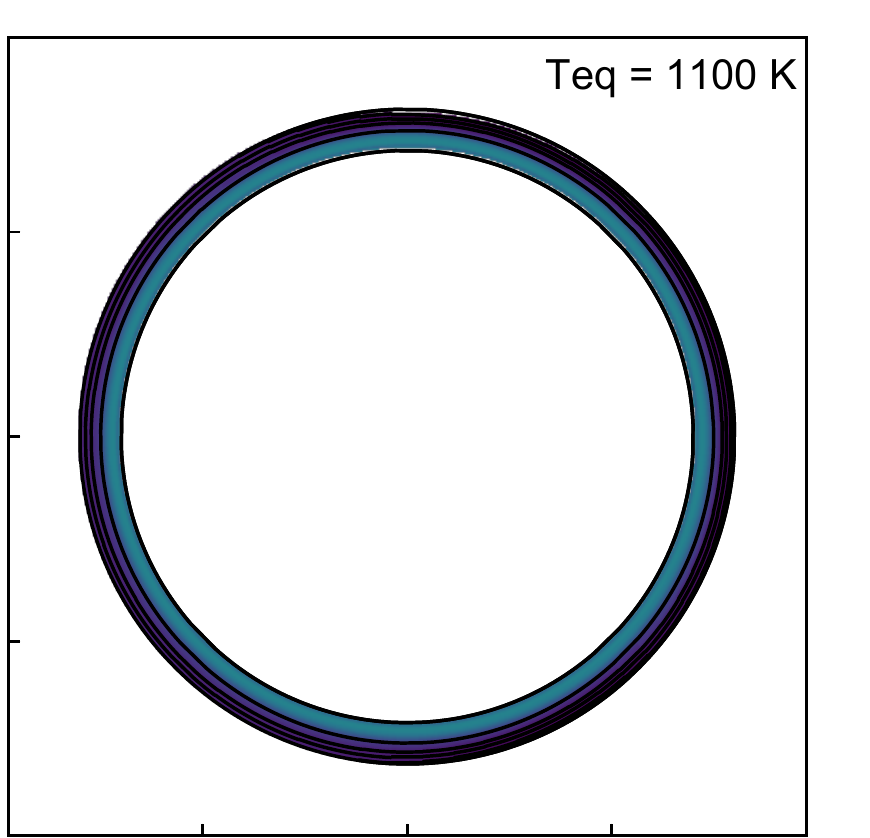}
\includegraphics[scale=\Tmap,trim = 3.9cm 1.65cm 2.5cm 0cm, clip]{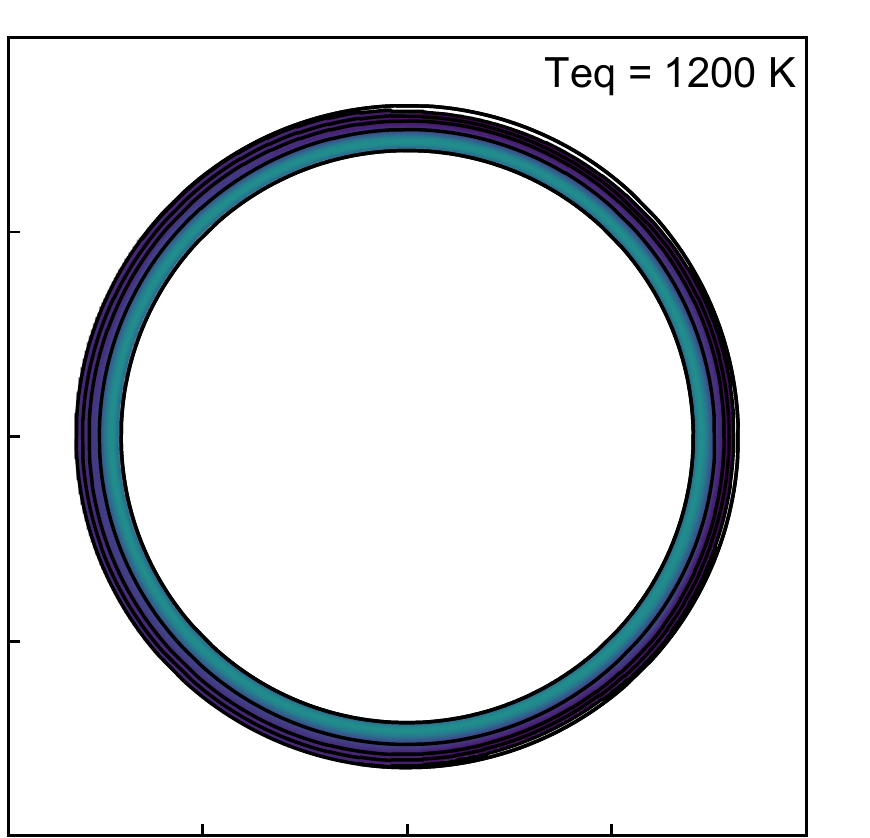}
\\
\includegraphics[scale=\Tmap,trim = 0.5cm 1.65cm 2.5cm 0cm, clip]{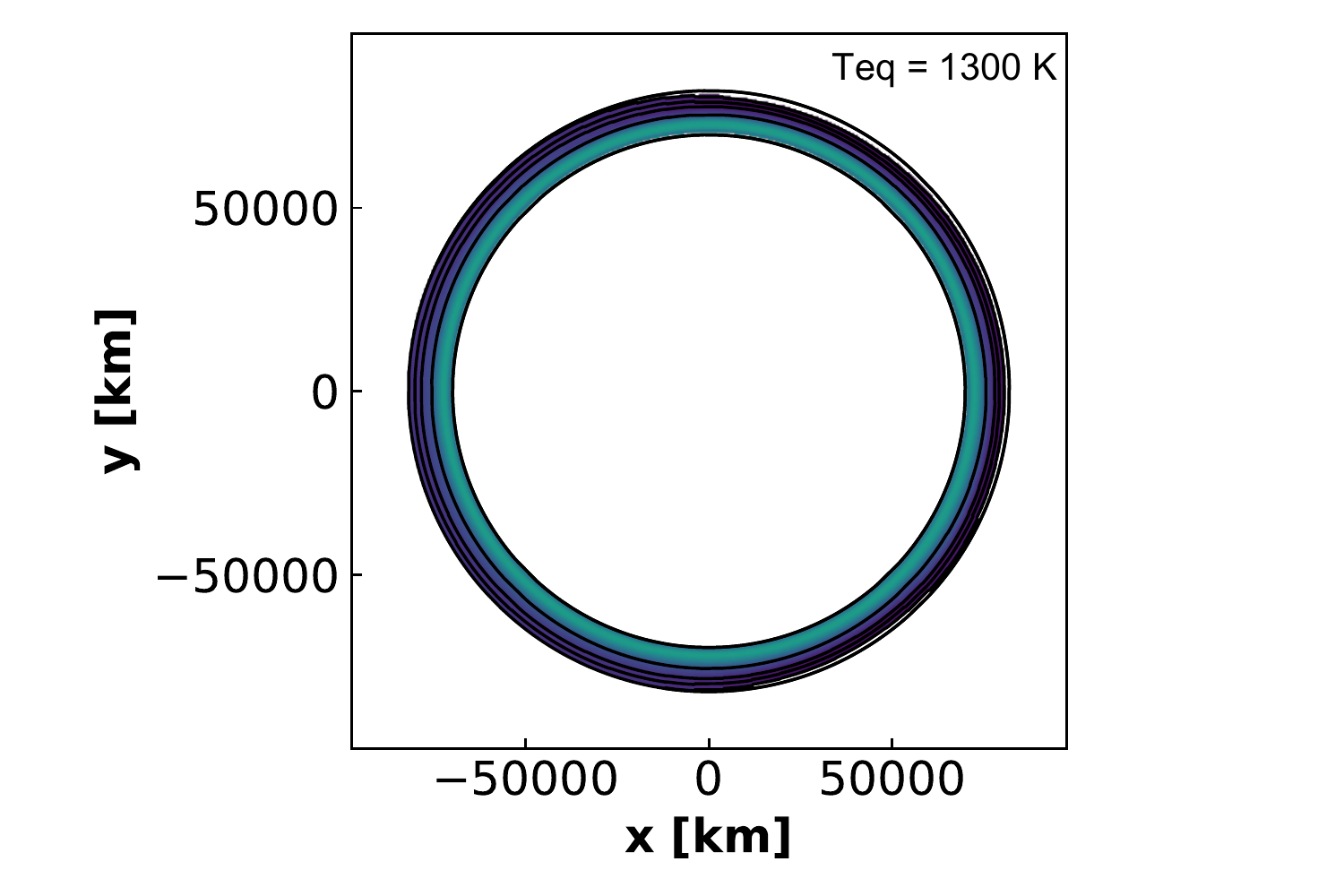}
\includegraphics[scale=\Tmap,trim = 3.9cm 1.65cm 2.5cm 0cm, clip]{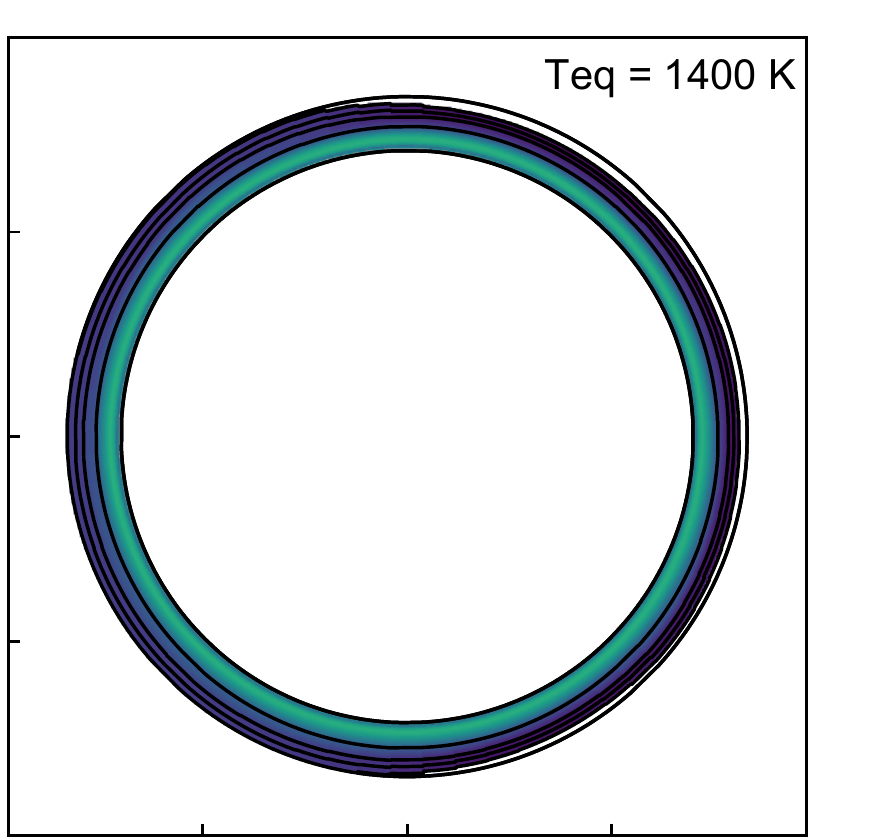}
\includegraphics[scale=\Tmap,trim = 3.9cm 1.65cm 2.5cm 0cm, clip]{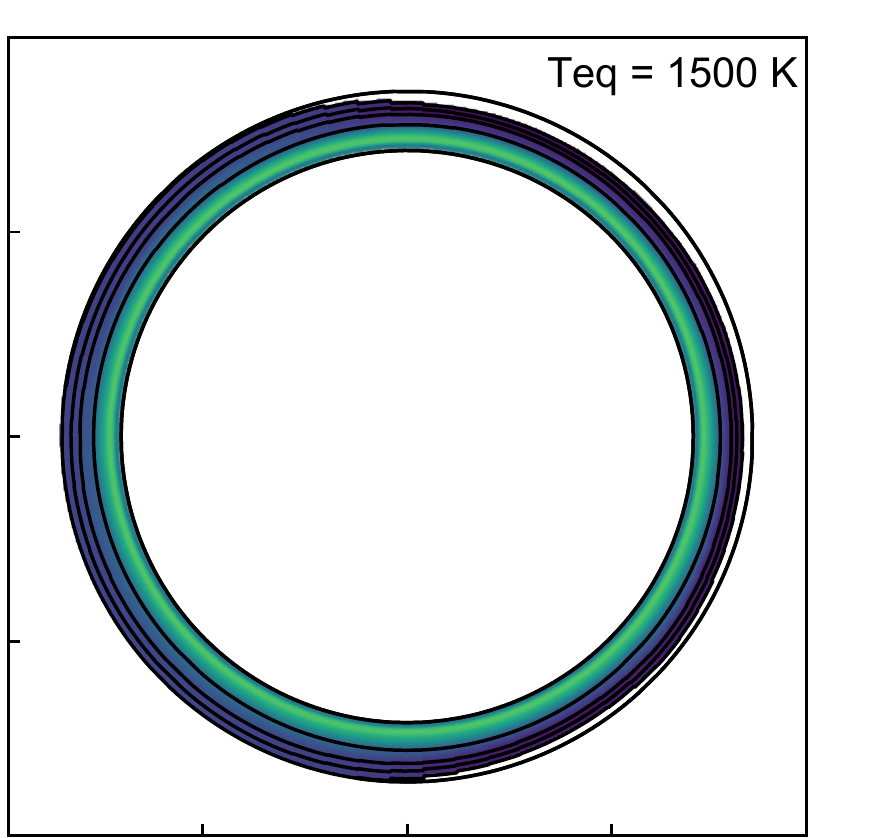}
\\
\includegraphics[scale=\Tmap,trim = 0.5cm 1.65cm 2.5cm 0cm, clip]{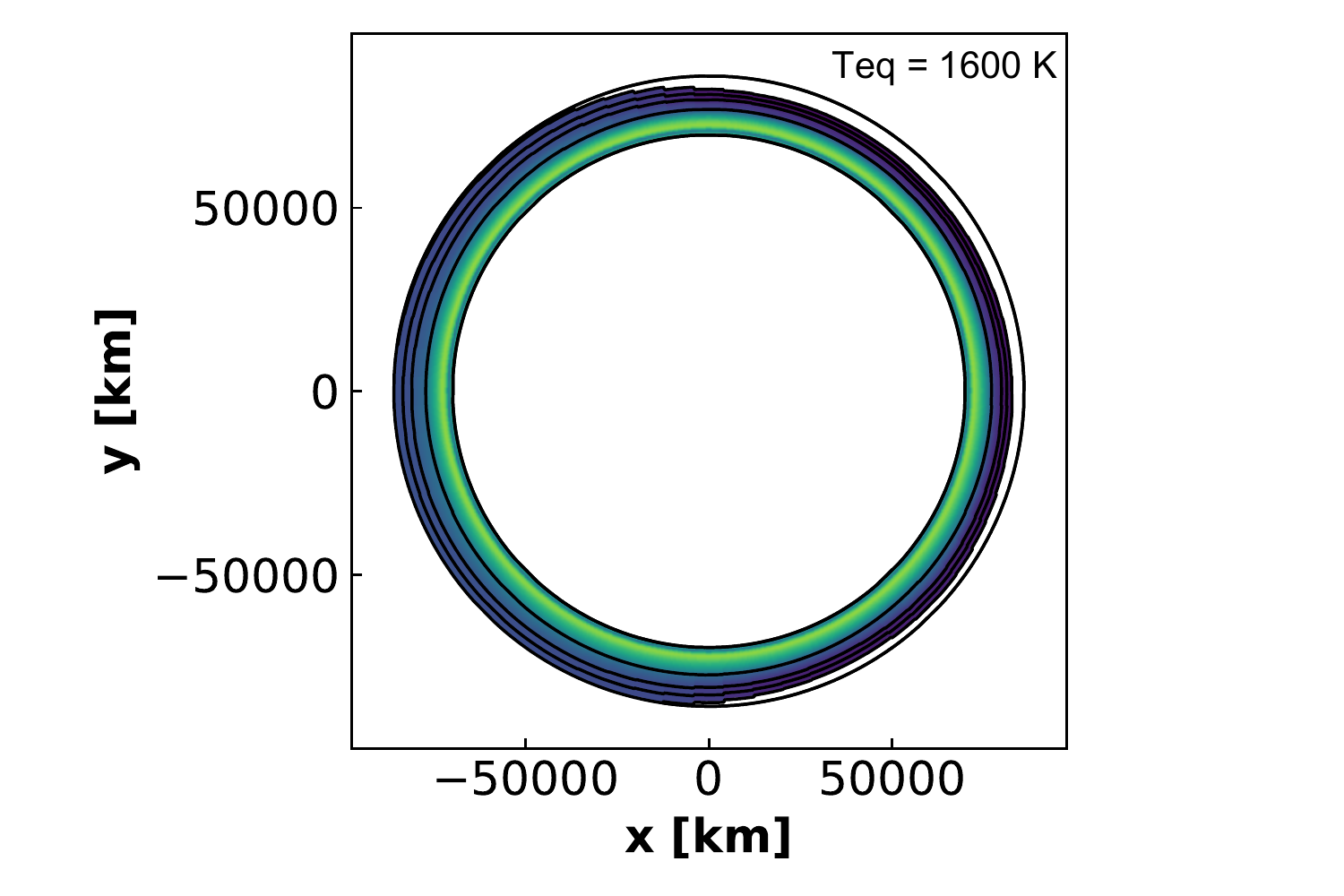}
\includegraphics[scale=\Tmap,trim = 3.9cm 1.65cm 2.5cm 0cm, clip]{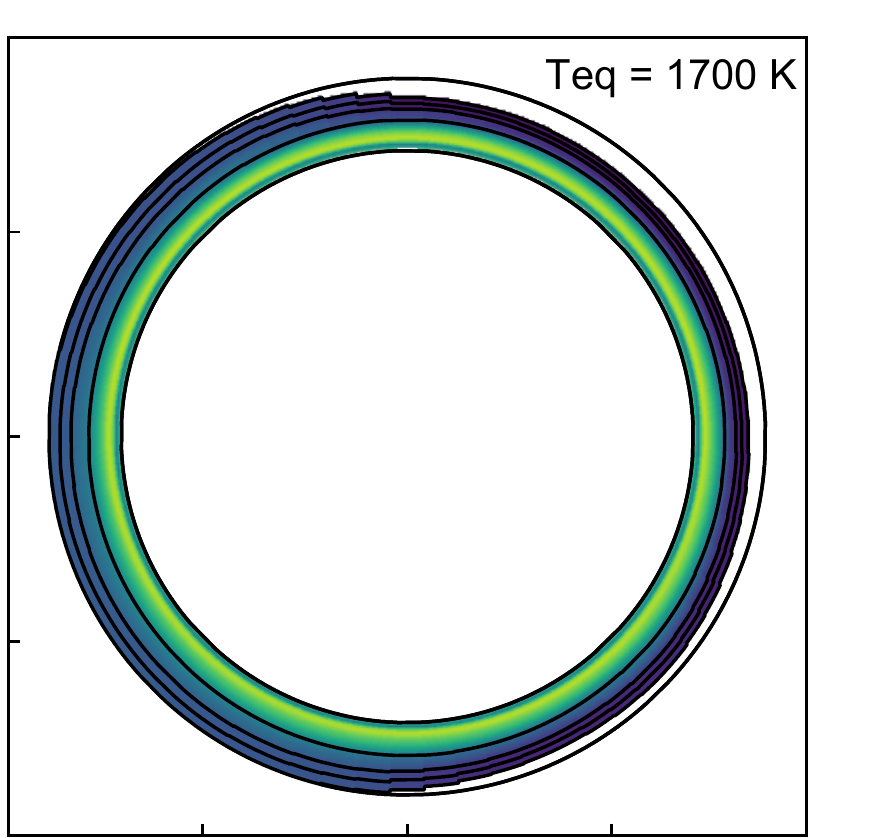}
\includegraphics[scale=\Tmap,trim = 3.9cm 1.65cm 2.5cm 0cm, clip]{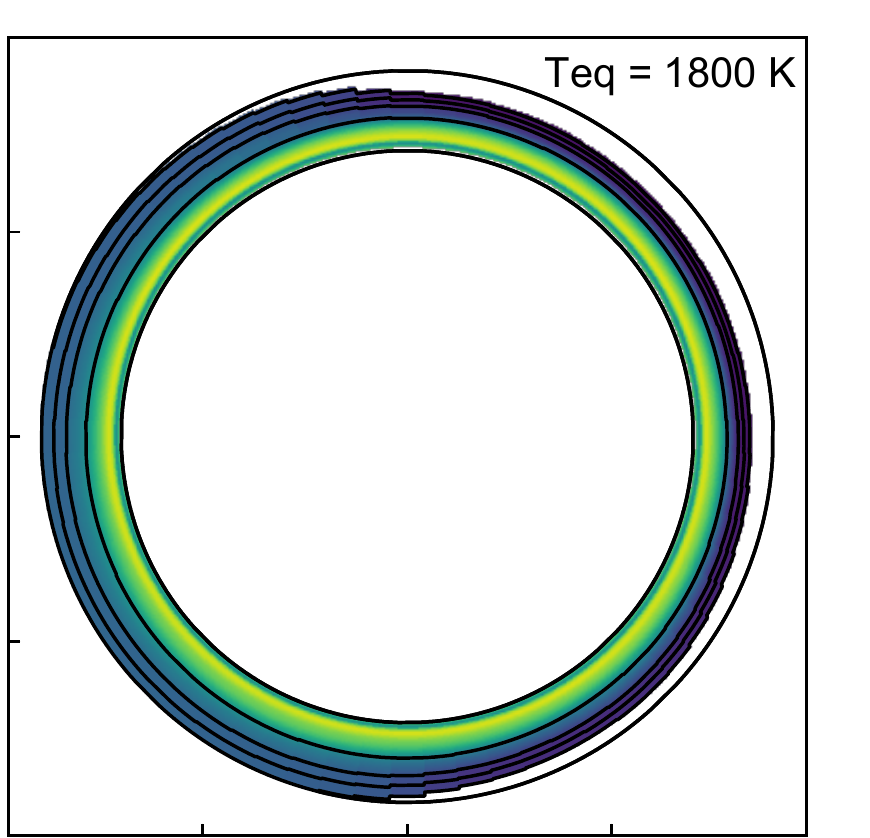}
\\
\includegraphics[scale=\Tmap,trim = 0.5cm 0cm 2.5cm 0cm, clip]{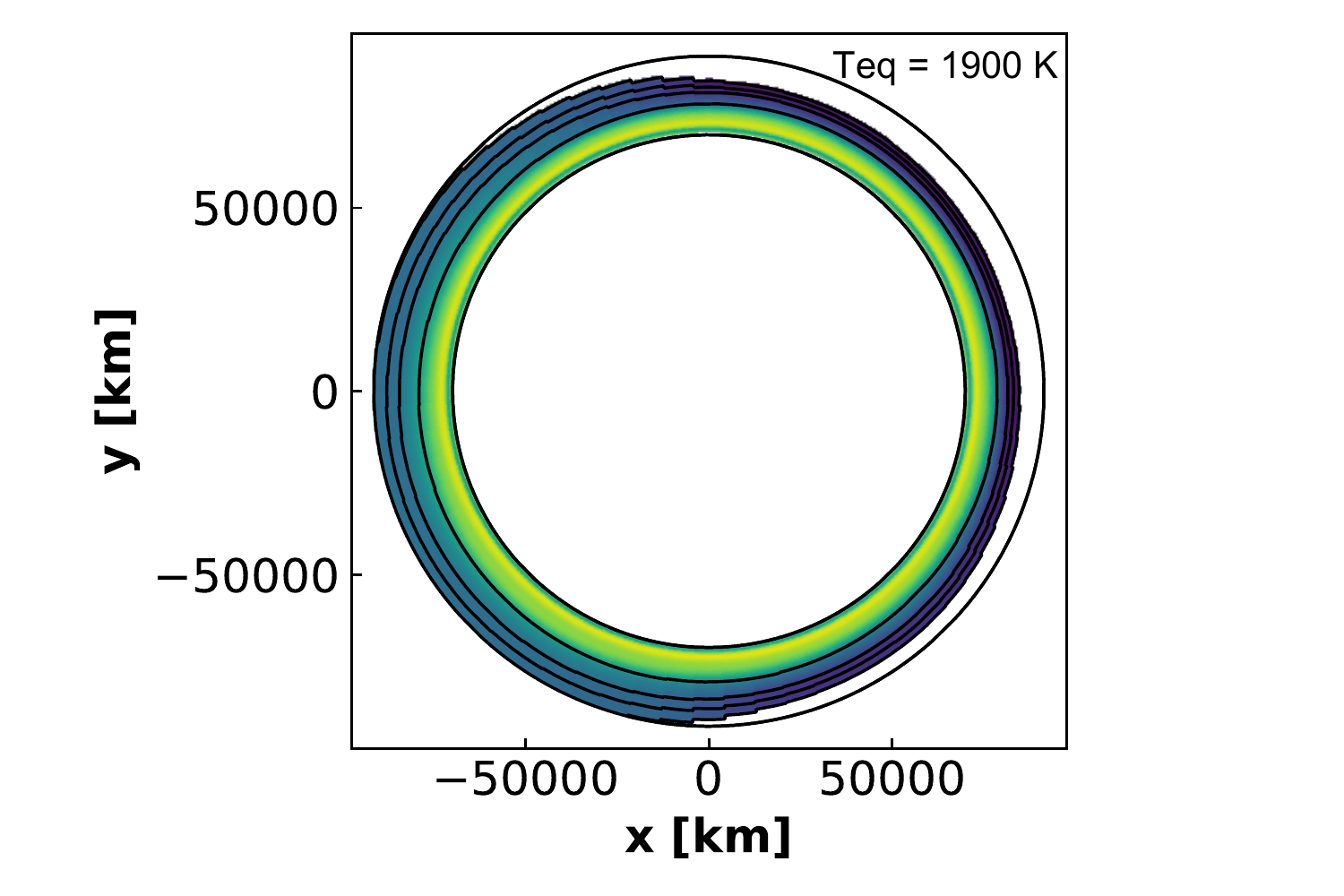}
\includegraphics[scale=\Tmap,trim = 3.9cm 0cm 2.5cm 0cm, clip]{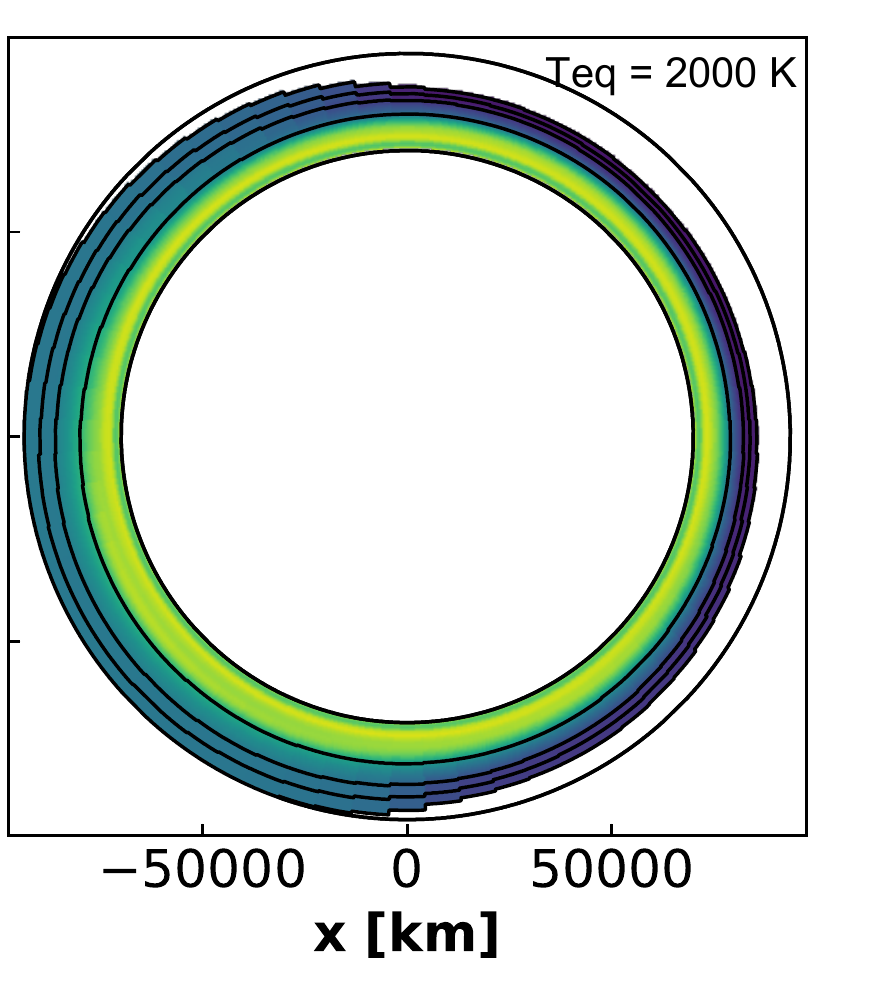}
\includegraphics[scale=\Tmap,trim = 3.9cm 0cm 2.5cm 0cm, clip]{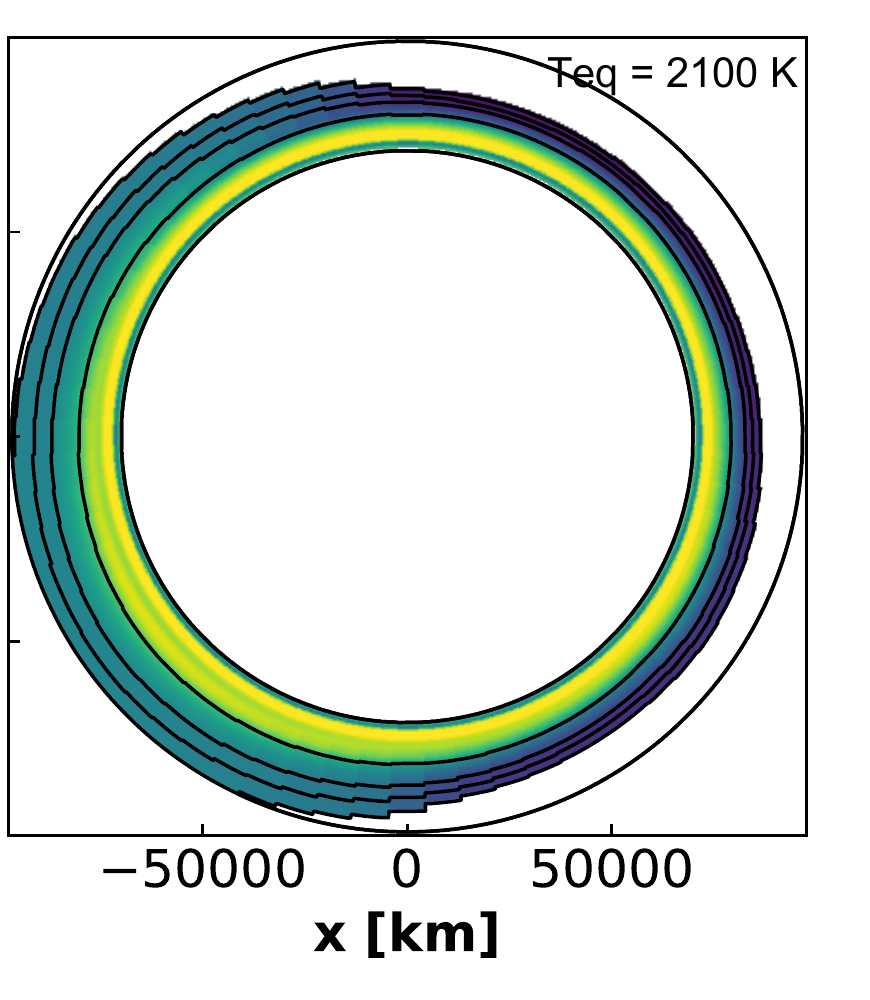}
\\
\includegraphics[scale=0.2,trim = 0cm 0cm 0cm 0cm, clip]{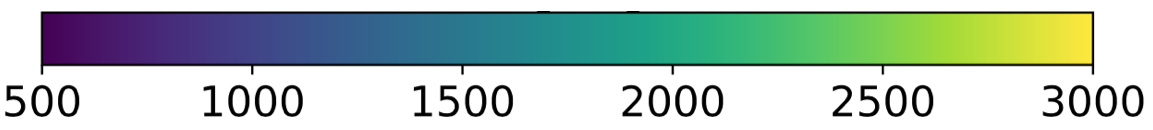}
\caption{Equatorial temperature maps for the 12 \textit{No thermal inversion} simulations performed without optical absorbers (VO and TiO), corresponding to 12 values of $T_\mathrm{eq}$.  
%in the atmosphere calculated respectively from top to bottom and from left to right for $T_\mathrm{eq}$ = 1000\,K until $T_\mathrm{eq}$ = 2100\,K with 100\,K step. 
The day-night dichotomy are shallower compared to Fig \ref{fig: maps-TiO} and the atmospheres extend less because of smaller scale heights. We also see a shift between the sub-stellar point and the hottest region of the atmosphere for the hottest planets (see Table \ref{tab: angle}).}
\label{fig: T-maps-noTiO}
\end{figure*}
Equatorial cut temperature maps for GCM simulations without TiO and VO in the atmosphere are presented in Fig.~\ref{fig: T-maps-noTiO}. These maps strongly differ from those obtained with TiO and VO. The atmospheres are less extended, especially in the day side due to the absence of the visible and near UV absorbers. Thus these atmospheres do not own stratospheres. In the absence of stratospheric heating, dayside scale heights are significantly smaller. The atmospheres are also much more homogeneous horizontally with no significant day-night dichotomy for $T_\mathrm{eq}$ below $\sim 1400$\,K. A pronounced day-night thermal gradient gradually appears for hotter cases with an increasing eastward shift of the hottest point with respect to the sub-stellar point that reaches 33$^{\circ}$ for the hottest simulation (see Table \ref{tab: angle}). The lower temperatures obtained without TiO and VO heating yield longer radiative timescale compared to the cases with thermal inversion, explaining these larger displacements of the hot spot.
%Here, the radiative time in the hottest atmospheres is greater than the dynamical time, which induces a shift of the hottest regions of the atmosphere to the east of the sub-stellar point. This situation is different from simulations with a hot stratosphere, where the very high temperature of the upper atmosphere on the day side decreases significantly the radiative timescale which becomes lower than the timescale of the advective processes.

The thermal profiles remain below the temperature required for a thermal dissociation of H$_2$ or H$_2$O, resulting in a total compositional homogeneity. Only a very particular region of the two hottest simulations allows a very low thermal dissociation of H$_2$O and H$_2$ associated with a decrease of H$_2$ and H$_2$O abundances by a factor of about 1.2. Moreover, as the concerned regions are contained between $50^\circ$ and $-50^\circ$ in latitude, $10^\circ$ and $70^\circ$ in longitude, and 2\,$10^{4}$ Pa and 5\,$10^{1}$ Pa in altitude, they are not probed in transmission. For this study, we can therefore consider that these atmospheres are chemically homogeneous.

\begin{figure}[h!]
\centering
\includegraphics[scale=0.28,trim = 0.3cm 0.2cm 1cm 0.8cm, clip]{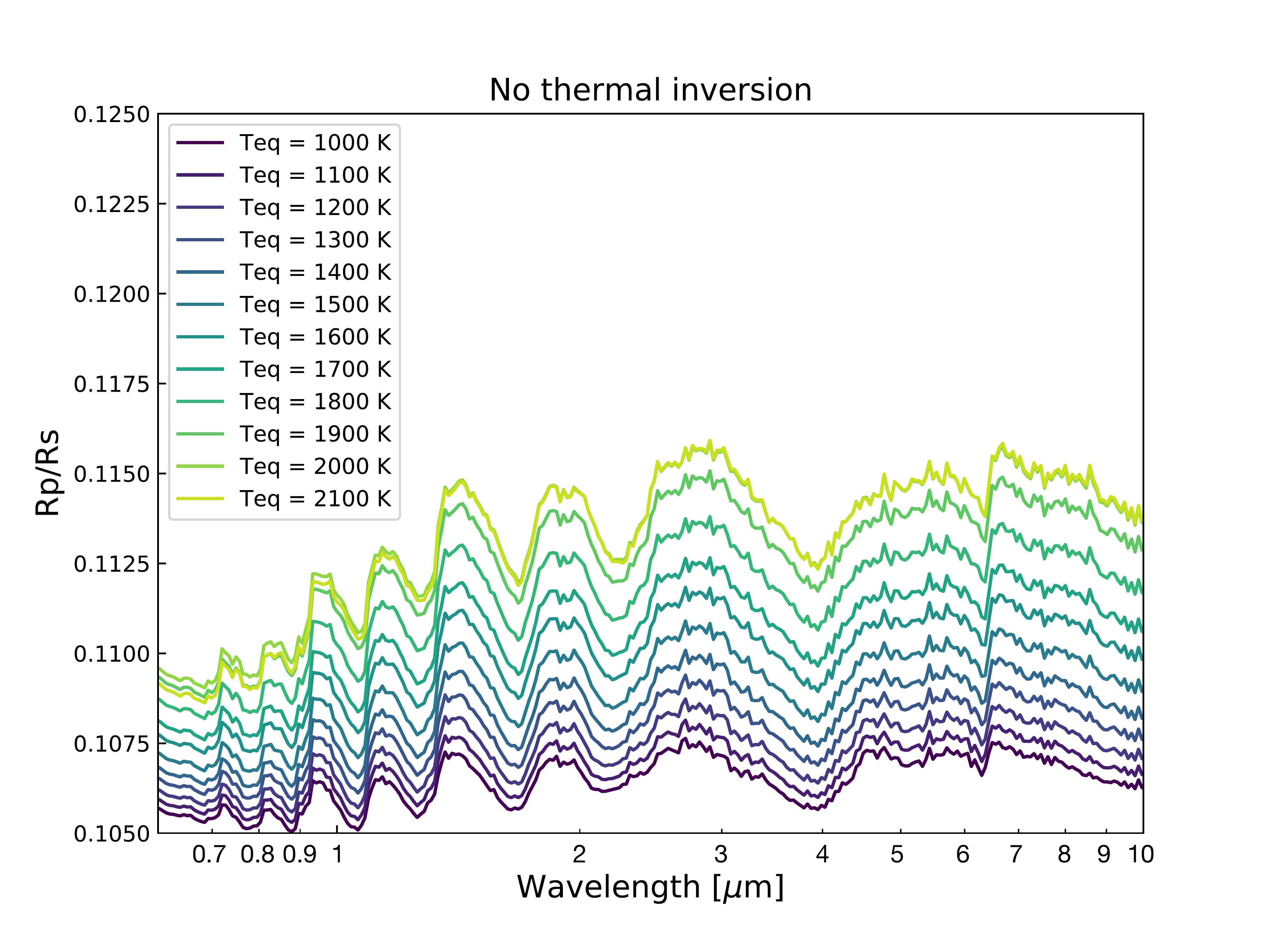}
\caption{Transmission spectra calculated with Pytmosph3R from the 12 simulations without optical absorbers (VO and TiO). 
%in the atmosphere, ranging from $T_\mathrm{eq}$\,=\,1000\,K to $T_\mathrm{eq}$\,=\,2100\,K in steps of 100\,K. 
As these simulations are overall colder without optical absorbers, the CO bands are hidden in the water bands and the apparent radii are smaller compared to the spectra in Fig~\ref{fig: Spectra-GCM}.}
\label{fig: Spectra-GCMnoTiO}
\end{figure}

The spectra generated by \pytmo for the 12 \textit{No thermal inversion} simulations are shown in Fig.~\ref{fig: Spectra-GCMnoTiO}.
If the spectra in transit for the coldest simulations (from $T_\mathrm{eq}$\,= 1000\,K to $T_\mathrm{eq}$\,= 1500\,K) are indeed very similar, the CO absorption bands for the warmer spectra are much less marked. The fact that H$_2$O does not dissociate in these simulations without thermal inversion implies that H$_2$O and CO contribute similarly to the transmission spectra. Thus, the CO bands, although present, do not stand out as clearly as when H$_2$O is dissociated and the spectrum probes these molecules at different temperatures (see Fig. \ref{fig: Spectra-GCMnoTiO}).

\begin{table}
\centering
\begin{tabular}{lll} 
\hline\hline
 & GCM w/ TI & GCM w/o TI \\
\hline
$T_\mathrm{eq}$ (K) & $\Gamma$ ($^\circ$) & $\Gamma$ ($^\circ$) \\
\hline 
1000 & - & $\sim$\,6 at 5.3\,10$^5$\,Pa \\
\hline
1100 & - & $\sim$\,6 at 5.3\,10$^5$\,Pa \\
\hline 
1200 & - & $\sim$\,6 at 5.3\,10$^5$\,Pa \\
\hline 
1300 & - & $\sim$\,6 at 5.3\,10$^5$\,Pa \\
\hline 
1400 & $\sim$\,23 at 11 Pa & $\sim$\,6 at 5.3\,10$^5$\,Pa \\
\hline 
1500 & $\sim$\,17 at 4 Pa & $\sim$\,6 at 5.3\,10$^5$\,Pa\\
\hline 
1600 & $\sim$\,11 at 8 Pa & $\sim$\,6 at 5.3\,10$^5$\,Pa\\
\hline 
1700 & $\sim$\,11 at 32 Pa & $\sim$\,11 at 5.3\,10$^5$\,Pa \\
\hline 
1800 & $\sim$\,6 at 44 Pa & $\sim$\,17 at 4.6\,10$^4$\,Pa\\
\hline 
1900 & $\sim$\,6 at 63 Pa & $\sim$\,23 at 2.3\,10$^4$\,Pa\\
\hline 
2000 & $\sim$\,0 at 2.6\,10$^2$\,Pa & $\sim$\,28 at 1.6\,10$^4$\,Pa \\
\hline
2100 & $\sim$\,0 at 2.6\,10$^2$\,Pa & $\sim$\,33 at 5.8\,10$^3$\,Pa\\
\hline\hline
\end{tabular}
\caption{Shift of the hot spot$^a$.}
\footnotesize{$^a$ $\Gamma$ angle (in degrees) between the sub-stellar point and the hottest point of the atmosphere with its pressure indicated for the GCM simulation with and without thermal inversion. The uncertainty on the angle is $\pm 3^\circ$ because of the resolution in longitude.}
\label{tab: angle}
\end{table}

We note that the $T_\mathrm{eq}$\,= 2100\,K spectrum is nearly superimposed with the $T_\mathrm{eq}$\,= 2000\,K,  spectrum above $1\:\muup$m and then very close to the $T_\mathrm{eq}$\,= 1900\,K below $1\:\muup$m. This spectra crossover is not found at lower values of  $T_\mathrm{eq}$, for which spectra are well distinct, overall shifted to higher apparent radii as $T_\mathrm{eq}$ increases. %the spectrum evolves differently from the other spectra, since it is almost superimposed on the spectrum of $T_\mathrm{eq}$\,= 2000\,K simulation's above 1\,$\mu$ m, then tends superimpose the spectrum related to the simulation at $T_\mathrm{eq}$\,= 1900\,K below 1\,$\mu$m. 
This could be explained by the strong east-west asymmetry of the limb produced by the zonal circulation also responsible for the hot-spot shift. The atmosphere is much more extended on the east limb compared to the west.   This trend is already visible in the simulations at $T_\mathrm{eq}$\,=\,1900\,K and $T_\mathrm{eq}$\,=\,2000\,K, but it is more dramatic for the simulation at $T_\mathrm{eq}$\,=\,2100\,K because the atmosphere shift is more intense here, as shown in Table \ref{tab: angle} which indicates the angle between the hottest point and the sub-stellar point of each simulation.
%\fsnote{I understand that this shift/asymmetry can be the explanation but it's not trivial to me why this effects have to stop the increase of R with Teq, and even reverse it and why the wavelength dependence. Indeed, the east limb is rising while the west one is shrinking and it's not obvious what the overall effect should be.} 

\section{Retrieval results}
\label{retrieval}

We performed 1D retrievals using \taurex on the set of transmission spectra previously described.

First, we computed retrievals assuming only \hho{} and CO as atmospheric trace gases in order to limit the number of free parameters, save computation time, and better identify biases. For each simulation, we retrieve 4 parameters: the planetary radius, H$_2$O and CO abundances, and a "gray" cloud layer. We also used the following two different TP profiles:
\begin{itemize}
\item[•] \textit{isothermal profile}: a single temperature is assumed for the whole atmosphere. This assumption is relevant for cold enough planets where only a thin part of atmosphere is probed;
\item[•] \textit{4-point thermal profile}: profile parameterized by 4 temperatures and 2 pressures. The top and bottom pressures are fixed at the extremes of the atmospheric model. This 1D vertical profile assumes a homogeneous atmosphere in latitude and longitude but introduces more freedom with a possible variation in altitude. This assumption is relevant when significant vertical variations are expected on the limb. With 6 parameters added, this profile costs more in computing time.
\end{itemize}

Then, we proceeded to a full retrieval analysis including TiO and VO abundances (2 more parameters), in addition to that of CO and \hho{}, as well as the 4-point thermal profile. Although the calculation took more time to converge, these tests allowed us to investigate the biases of a more complex atmosphere and to determine whether TiO and VO spectral features better constrain the retrieval or not.

Note that for every retrieval done in this paper, the retrieved abundances are constant for the whole atmosphere and the \Heratio\ ratio is set at the solar metallicity. In addition, \taurex\ does not take into account the thermal dissociation of species.

\subsection{Isothermal retrievals}
\label{isothermal}

\begin{figure*}[h!]
\centering
\includegraphics[scale=0.7]{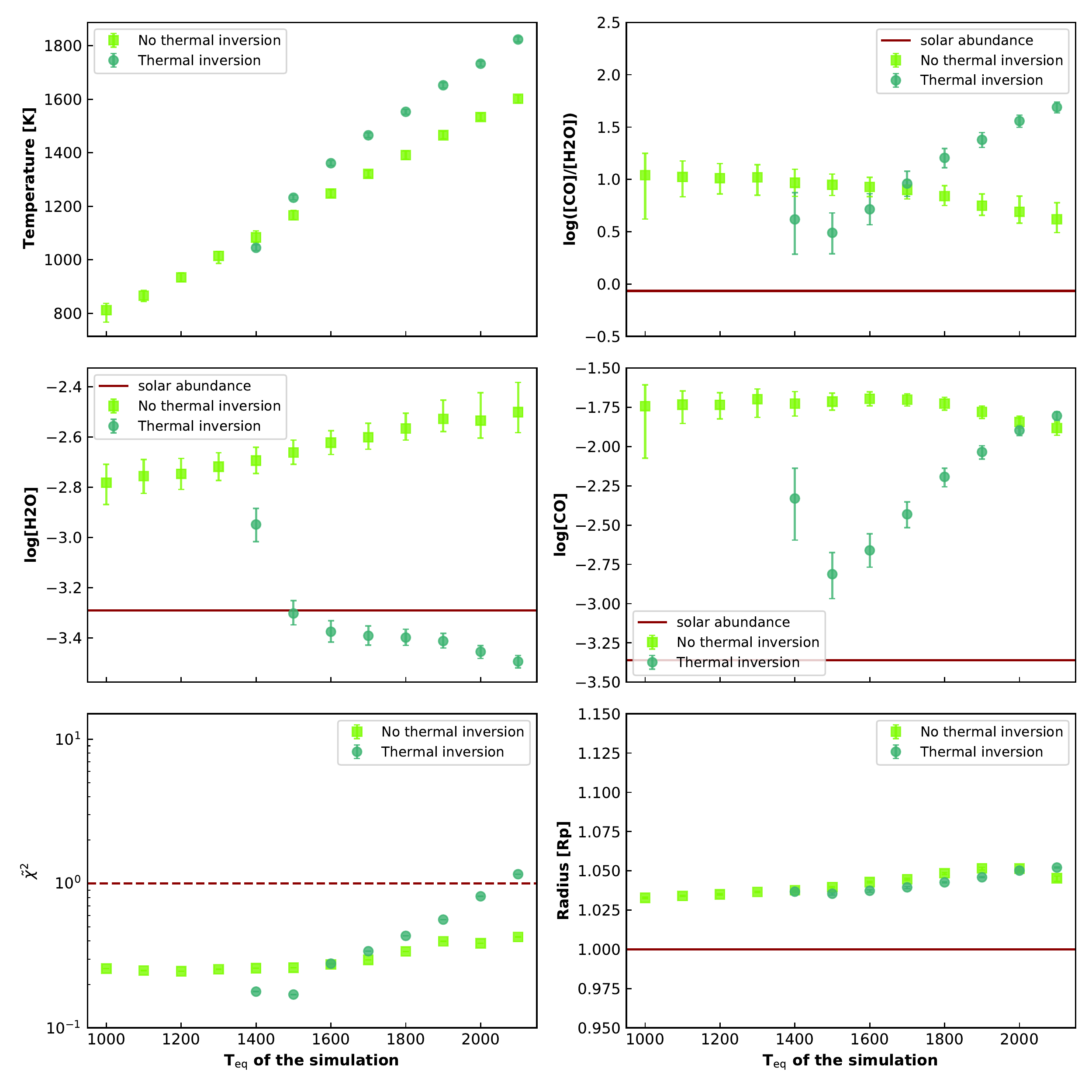}
\caption{Retrieval results with \taurex assuming an isothermal profile. Cases with and without thermal inversion in the input atmosphere are shown in light and dark green, respectively. We plot the temperature (top left), \COratio (top right), the CO (middle right) and the \hho{} (middle left) abundances, the planetary radius (bottom right) within 1-$\sigma$ error, and $\redchid$ (bottom left). These results have been obtained with non-randomized spectra but $\redchid$ calculations and \taurex assume a 30~ppm 1-$\sigma$ Gaussian noise on the whole spectral domain. The red line represents the input value in the GCM simulations and the red dashed line shows where $\redchid = 1$.}
\label{fig: comp-plot-iso}
\end{figure*}

The isothermal profile is commonly used in retrieval analyses of transmission spectra based on the assumption that only a small and homogeneous region around the terminator is probed \citep{Evans_2018,Tsiaras2018,Edwards-ares2020}. Moreover, its reduced number of free parameters is consistent with the weak information content of most low-resolution small-bandwidth available spectra.  This assumption is often well justified but deserves to be tested in the context of forthcoming higher quality spectra that JWST and Ariel will deliver.

The retrieval results obtained with \taurex using the isothermal assumption are shown in Fig.~\ref{fig: comp-plot-iso}. 
For each simulation, we show the temperature, the planetary radius, and the \hho\ and CO abundances retrieved. We also show the CO over \hho\ abundance ratio, which indicates a departure from the solar composition. As explained in Sec.~\ref{noise analysis}, we use non-randomized spectra and we indicate the $\redchid$ that is computed with the best-fit parameter values derived from the posterior distribution given by \taurex. We stress i) that with non-randomized spectra, $\redchid \ll 1$ is an acceptable value, not the sign of noise fitting, and ii) that model comparison is done with logarithmic Bayes factors. 
%
%As we described in Sec. \ref{num exp}, we used nonrandomized spectra to avoid adding more parameters biases and prevent dealing with the degeneracy of the solution \citep{Chiavassa2017}. To check the statistical significance of the fit given by \taurex, we calculate the reduced-\redchi\ using the best fit given by \taurex (see Eq. \ref{redchi}) where $O$ and $C$ are respectively the observed and calculated spectra, $\sigma$ are the uncertainties, $p$ the number of parameters and $N$ the total number of points. As we have nonrandomized spectra, the reduced-\redchi\ can be lower than 1 which indicates that the fit cannot be statistically rejected when it happens. Thus, as soon as the reduced-\redchi\ is less than 1, we consider the retrieval valid.

\taurex always converges to a single solution but the goodness of the fit will differ if TiO and VO are present or not. In the case of simulations without TiO and VO, \taurex always produces a good fit ($\redchid{} \ll 1$), with a notable slight increase in $\redchid$ for the hottest simulations. The outcome is different for simulations with TiO and VO. For the coldest simulations, therefore with little thermal dissociation in the atmosphere (see Fig. \ref{fig: maps-TiO}), $\redchid$ is less than 1. This is consistent with simulations without TiO and VO in the atmosphere where no thermal dissociation takes place. In contrast, $\redchid$ increases very clearly with increasing equilibrium temperature, exceeding 1 for all the other simulations. This behavior is similar to that shown by \citet{Pluriel2020}, indicating that \taurex fails to correctly take the thermal dissociation of \hh\ into account.

The results show that none of the retrieved abundances are in agreement with their actual values in the simulations. As a reminder, we simulate atmospheres with solar abundances and where CO does not dissociate. Therefore, we would expect to retrieve solar abundances for CO, and abundances smaller or equal to the solar abundance for \hho{} (because \hho{} can dissociate).

\begin{figure}
\centering
\includegraphics[scale=0.2]{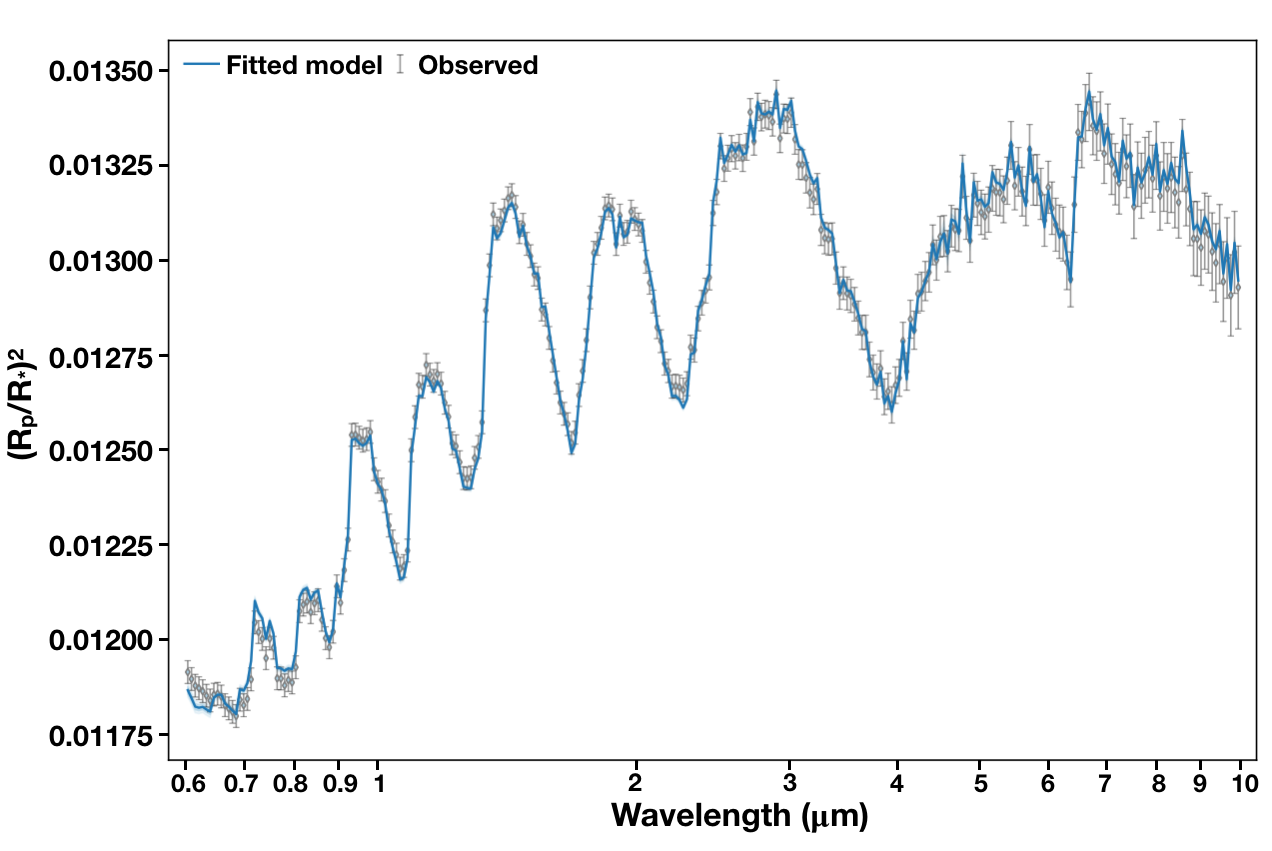}
\caption{Best fit retrieved spectrum using an isothermal profile parametrization (in blue). The input spectrum has been calculated with Pytmosph3R for the $T_\mathrm{eq}$\,=\,2100\,K simulation without optical absorbers (VO and TiO) in the atmosphere (in grey).}
\label{fig: Spectrum-fit}
\end{figure}

The retrieval results from the simulations without TiO and VO in the atmosphere are wrong. The retrieved \COratio is equal to about 10 and the retrieved abundances of CO and \hho\ are always above the solar abundance. This indicates that \taurex is not able to retrieve results consistent with the \textit{ground truth}, even though we have a visually good fit and $\redchid < 1$, all of which would give an undue confidence to an observer. Fig.~\ref{fig: Spectrum-fit} shows an example of a good agreement between the data and the retrieval (for the simulation at $T_\mathrm{eq}$\,=\,2100\,K) which leads to wrong parameters value. These results are unexpected because, as shown in Fig.~\ref{fig: T-maps-noTiO}, temperature never reaches a value high enough so that species such as \hho{} or \hh would dissociate. Therefore, we do not expect to observe strong compositional heterogeneities between the day side and the night side, and therefore we would expect to retrieve solar abundances. We will see in Sec.~\ref{vertical profile} that the assumption of an isothermal profile is the main culprit for retrieving wrong parameter values.

Retrievals are more consistent in the case of \textit{Thermal Inversion} simulations including TiO and VO. For the coldest simulations, the retrieved \hho{} abundance is almost solar even if it slightly decreases from the solar abundance for the hottest simulations. The CO abundance starts to be well constrained in the coldest simulation then it becomes more and more biased with the hottest simulations. Thus, we observe an increase in \COratio when we simulate hotter Jupiters. This behavior indicates that the 1D retrieval models are less biased when applied to cooler atmospheres. The difference between the scale height on the day and on the night is reduced when the temperature drops, therefore the difference in altitude of the CO probed on the day side compared to the \hho{} probed on the night side decreases. On the other hand, the amount of dissociated \hho{} also decreases for the coldest planets, implying that the atmosphere is probed less deeply and at higher temperatures, hence a smaller difference in the spectra appears in Fig.~\ref{fig: Spectra-GCM}.

We observe a regular increase in the retrieved temperature with the equilibrium temperature of the planet for simulations without TiO and VO in the atmosphere, which is expected as the atmosphere are globally hotter (see Fig. \ref{fig: T-maps-noTiO}).
%The retrieved temperatures are consistent with the simulations. We observe a regular increase in the retrieved temperature with the equilibrium temperature of the planet for simulations without TiO and VO in the atmosphere, which is consistent with the temperature maps of GCM simulations (see Fig. \ref{fig: T-maps-noTiO}). 
For simulations with TiO and VO in the atmosphere, the increase in temperature is regular until a break in slope from $T_\mathrm{eq}$\,=\,1500\,K. This behavior can be explained by the shape of the temperature maps in Fig. \ref{fig: maps-TiO}, where the east-west asymmetry becomes less important for the hottest cases. \taurex seeks here to best fit the very strong absorption bands of CO, and does so by increasing the abundance of the species, so as the retrieved temperature.

The retrieved radius is almost constant for all the simulations, quite close to the input radius of the simulations. It is therefore difficult to extract much information from this retrieved parameter which does not seem to play an important role here in improving the fit of \taurex retrievals.

To summarize this section, isothermal retrievals are insufficient to get the complexity of hot exoplanetary atmospheres, even for the more homogeneous simulations such as the one without optical absorbers. The discrepancy between the retrieved parameters and the \textit{ground truth} is usually considerable, much larger than uncertainties estimated from posterior distributions and regardless the goodness of the fit. The following step is thus to check whether assuming a more complex vertical thermal profile could solve this issue.

\subsection{4-point TP profile retrievals}
\label{vertical profile}

We present here the results of the retrieval procedure that no longer assumes an isothermal vertical profile but a 6-parameter thermal profile (Sec.~\ref{retrieval}). 
%The retrieved values for all simulations are presented in Fig.~\ref{fig: comp-plot-4pts}. They are more complex retrieval based on the \textit{npoint} profile describe sooner in Sec \ref{retrieval}.

\begin{figure*}
\centering
\includegraphics[scale=0.7]{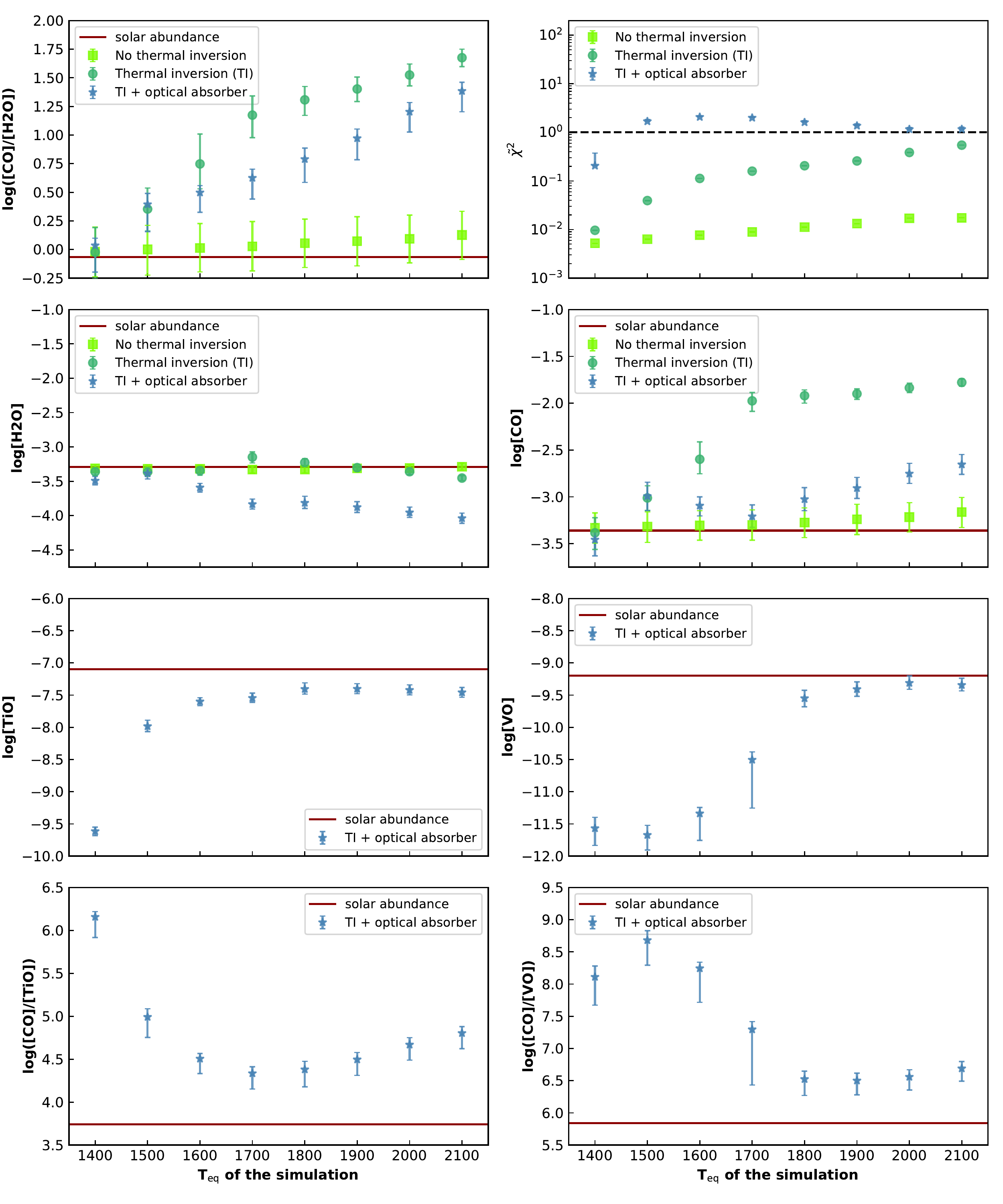}
\caption{Retrieval results with \taurex using a 4-point parametrization for the vertical thermal profile. We present 4 set of retrievals. The first one corresponds to the analysis of spectra obtained from the \text{No Thermal Inversion} simulations  (light green squares). Then we have 3 different ways to analyze the spectra from the \text{Thermal Inversion} simulations. The dark green circles show retrievals ignoring TiO and VO, then the light blue triangles and dark blue stars shows retrievals considering TiO and VO without and with condensation, respectively. We plot \COratio,  \TiOratio, \VOratio, and the CO, the \hho{}, the TiO, the VO abundances within 1-$\sigma$ errors and $\redchid$. These results have been obtained with non-randomized spectra, but $\redchid$ calculations and \taurex assumes a 30~ppm 1-$\sigma$ Gaussian noise on the whole spectral domain.  The red line represents the input value in the GCM simulations and the red dashed line shows where $\redchid$ = 1.}
\label{fig: comp-plot-4pts}
\end{figure*}

%\begin{figure*}
%\centering
%\includegraphics[scale=0.85]{T-P_profiles_comparison_noTiO}
%\caption{Temperature profiles of GCM models without TiO/VO in the atmosphere for $T_\mathrm{eq}$\,=\,1400\,K (left) and $T_\mathrm{eq}$\,=\,2100\,K (right). From top to bottom, we have the east limb (longitude 90$^\circ$) and the west limb (longitude 170$^\circ$) for every latitudes (color gradient from blue to red, respectively from the north pole to the south pole). We plot also the equator (latitude 0$^\circ$) for every longitudes (color gradient from blue to red, respectively from the western limb to the eastern limb). We added the retrieved temperature profile in the isothermal mode (grey) and in 4point PT profile mode (black).}
%\label{fig: T-P_noTiO}
%\end{figure*}

\subsubsection{Without thermal inversion}
\label{no inversion}

\begin{figure}
\centering
\includegraphics[scale=0.6,trim = 0cm 3cm 0cm 3cm,clip]{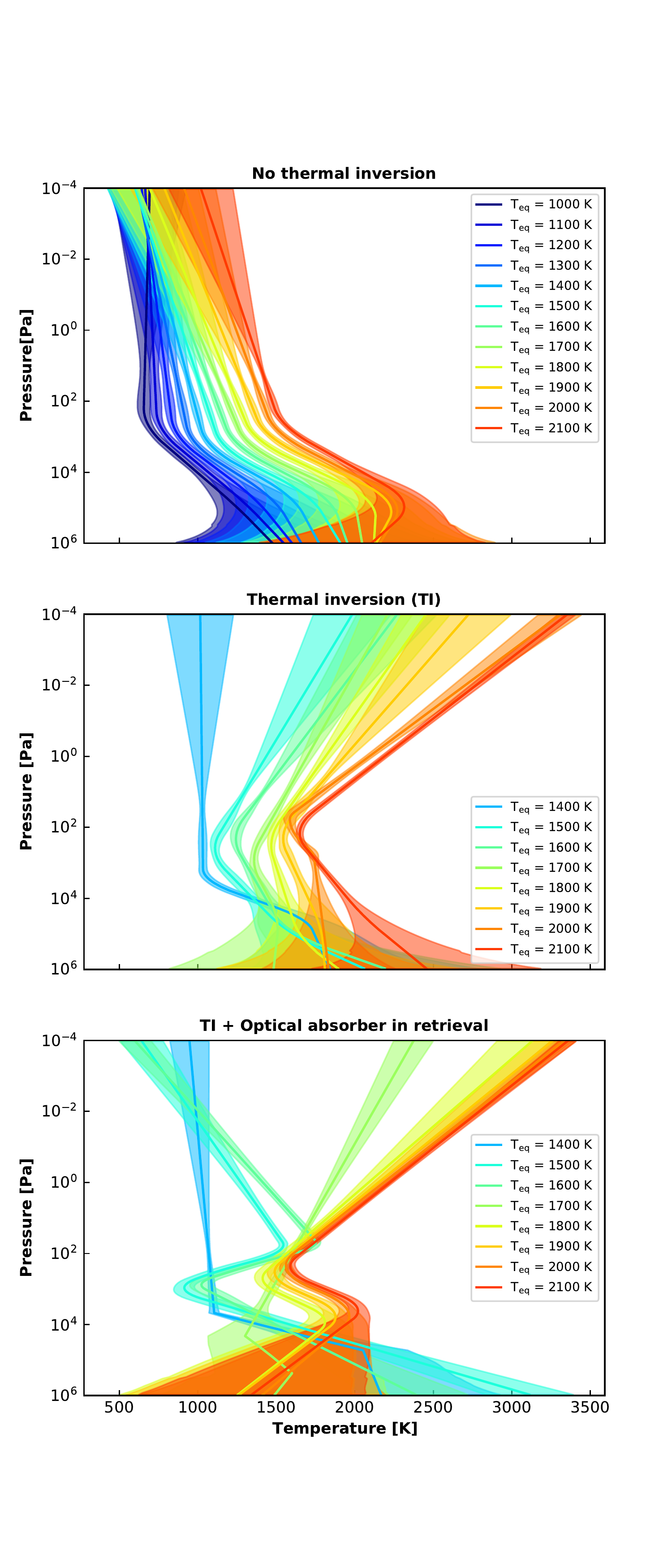}
\caption{Retrieved temperature profiles for the GCM simulation: (top) without thermal inversion; (middle) with thermal inversion assuming a simplified atmosphere; (bottom) with thermal inversion assuming a more complex atmosphere with optical absorber (TiO and VO) in the atmosphere.}
\label{fig: T-P_4pts}
\end{figure}

We first focus on the input simulations (Fig.~\ref{flowchart}) without optical absorbers (TiO and VO) in the atmosphere, and therefore, without thermal inversion. The retrieval results are here consistent with the input models. The retrieved abundances of CO and \hho{}, consequently \COratio, are now well constrained and fit the solar abundance within 1-$\sigma$ in every retrieval. We see here that vertical effects are not negligible to retrieve correclty such atmospheres, because isothermal retrievals were biased. Thus, we need to let \taurex fit his own vertical profile. The thermal profiles retrieved within 1-$\sigma$ are shown in Fig.~\ref{fig: T-P_4pts} (top). We use a log-linear interpolation between the temperature nodes. To quantify the improvement of retrieving a 4-point thermal profile, compared to the isothermal assumption, we calculate the logarithmic Bayes factor following Eq.~\ref{eq:logbayes_factor}.

\begin{figure}
\centering
\includegraphics[scale=0.55]{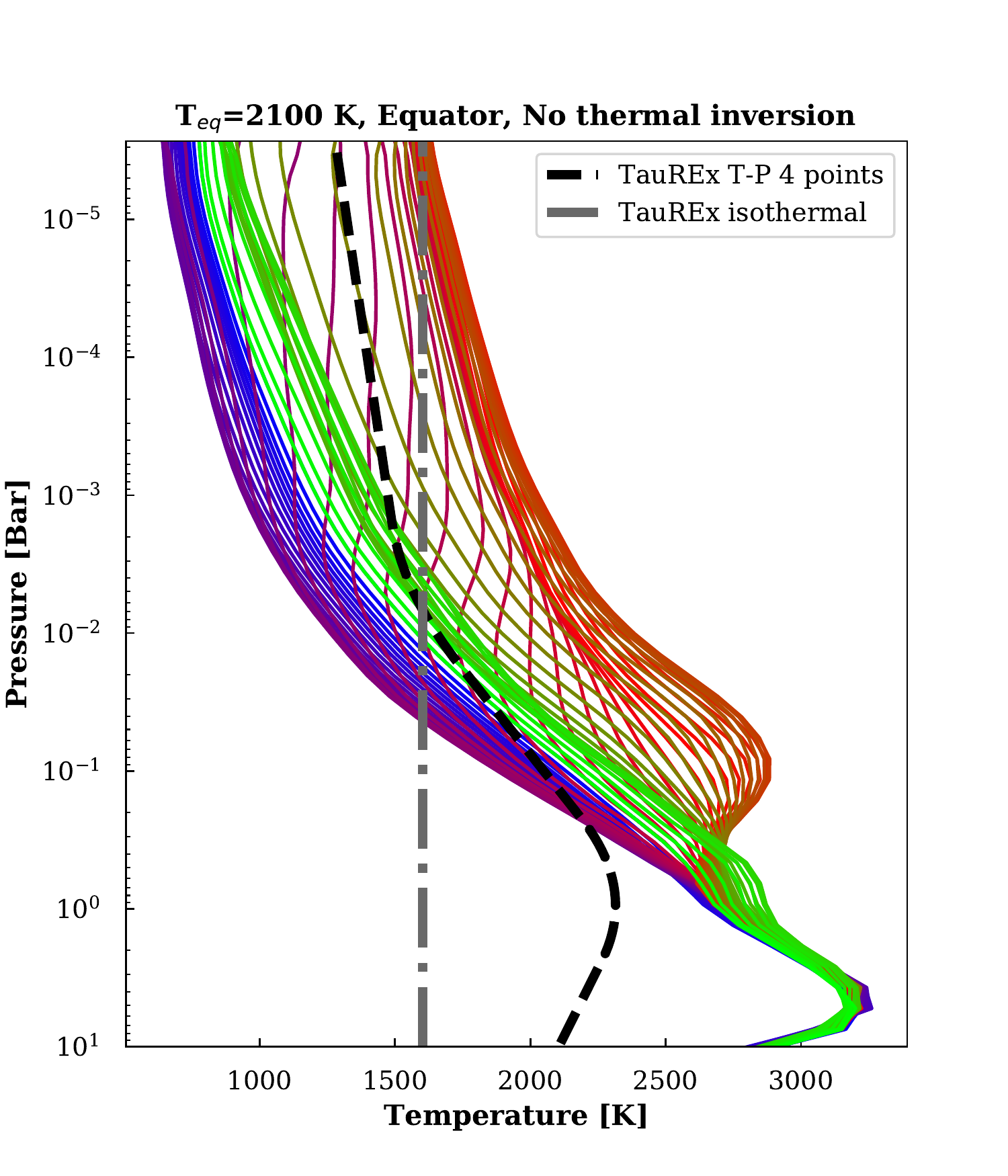}
\includegraphics[scale=0.35,trim = 0cm 0cm 7.5cm 0cm,clip]{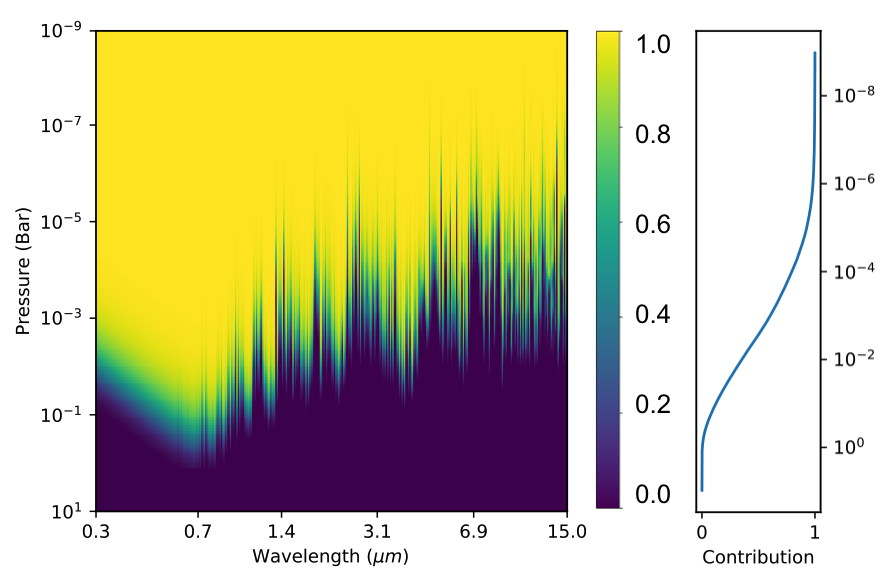}
\caption{\textbf{Top:} Equatorial thermal profiles from the \textit{No Thermal Inversion} $T_\mathrm{eq}$\,=\,2100\,K simulation. Each color represents a longitude from anti-stellar (blue) to sub-stellar (red). Profiles near the terminator and probed by transmission are green. The dotted-line grey (isothermal) and dashed-line black are the retrieved profiles. \textbf{Bottom:} Transmittance map (colorbar from 0 to 1) for each wavelength at each pressure.}
\label{fig: contrib-noTiO}
\end{figure}

We see in Table~\ref{tab: bayes-factor} that \taurex strongly favors the 4-point thermal profile compared to the isothermal one. To understand why vertical effects are important, we show in Fig.~\ref{fig: contrib-noTiO} the contribution function of the transmission spectra for the $T_\mathrm{eq} = 2100$~K simulation and the thermal profiles both retrieved and from the input GCM. We only plotted the GCM profile around the limb to focus on the probed region. The contribution function shows that the regions probed covers 6 orders magnitude in pressure, from around $10^{5}$ to $10^{-1}$~Pa. Depending on the wavelength, the features in the transmission spectra are therefore coming from regions at different temperatures. 

An isothermal profile does not manage do describe this complexity, especially where a broad range of pressure is probed. Fig.~\ref{fig: contrib-noTiO} shows the thermal profile at the equator for every longitude in the hottest simulation. The blue and the red curves represent respectively the anti-stellar and the sub-stellar regions of the atmosphere. The regions probed in transmission are thus mainly represented by the green curves. We also show the isothermal and vertical retrieved thermal profiles in Fig.~\ref{fig: contrib-noTiO}. The TP profiles cannot be well approximated by an isothermal profile because the input profiles moves away from an isothermal profile by several hundreds of kelvins when the vertical retrieved TP profile fits better the GCM TP profile probed. Though, as it is shown in the contribution plot at the bottom of Fig.~\ref{fig: contrib-noTiO}, regions from $\sim 10^{4}$~Pa (around 0.7~µm) to $\sim 10^{-1}$~Pa are probed. From the shape of the thermal profiles in the terminator region and from the large pressure range probed by the transmission spectrum, we can conclude that an isothermal model is not well-suited and will either fail to yield a correct fit or will yield a correct fit with parameters significantly departing from the ground truth to compensate for this limitation. We clearly fall in the second case here as \taurex has to increase the species abundances well over their actual values to match the observed features, which is no longer needed when we allow the profile to be non-isothermal.

For this reason, the model using a vertical profile manages to retrieve the input abundances as shown in Fig.~\ref{fig: comp-plot-4pts} with a higher level of confidence than in an isothermal retrieval as it is shown in Table~\ref{tab: bayes-factor}. This is an important result since it clearly indicates that the assumptions of 1D retrieval models are justified to analyze and interpret the observations obtained on planets that are not too hot. However it confirms the caveat about the isothermal assumption (see Sec. \ref{isothermal}) which leads to wrong parameters value despite an excellent agreement ($\redchid\,\leq\,1$). Of course, this is a simplified model and we could well imagine the presence of a species on one side of the atmosphere and not on the other which could bias the results even assuming a non-isothermal vertical TP profile.

%\fsnote{What strikes me is the fact that in the case of real observation without any independant knowledge of the ground truth, one would have to use both models (isothermal and n-points profiles) and compare to conclude that, despite an excellent agreement, the isothermal model is crap. Which means that studies that use only one model - and most studies do that - have absolutely no argument to defend the robustness of their inferrance. I don't know if what I say here is trivial or not...}

\begin{table}
    \centering
    \begin{tabular}{ccc}
    \hline \hline
    \multicolumn{3}{c}{Bayes factor (vertical vs isothermal TP profiles)}\\
    \hline
    $T_\mathrm{eq}$ (K) & w/o thermal inversion & w/ thermal inversion \\
    \hline
    1000 & 32.10 & - \\
    \hline
    1100 & 31.62 & - \\
    \hline
    1200 & 31.26 & - \\
    \hline
    1300 & 33.14 & - \\
    \hline
    1400 & 33.26 & 20.06 \\
    \hline
    1500 & 33.98 & 15.17 \\
    \hline
    1600 & 35.55 & 19.82 \\
    \hline
    1700 & 38.14 & 22.72  \\
    \hline
    1800 & 43.10 & 27.43 \\
    \hline
    1900 & 51.60 & 32.82 \\
    \hline
    2000 & 50.13 & 50.25 \\
    \hline
    2100 & 55.30 & 77.74 \\
    \hline\hline
    \end{tabular}
    \caption{Logarithmic Bayes factor$^a$.}
    \footnotesize{$^a$Logarithmic Bayes factor comparing retrievals done assuming an isothermal or non-isothermal vertical profile for both the \textit{Thermal Inversion} and \textit{No Thermal Inversion} simulations. Bayes factor for non-isothermal profiles are 15 to 78 times larger than those obtained with isothermal profiles, suggesting a very strong preference for the non-isothermal model.}
    \label{tab: bayes-factor}
\end{table}

%\begin{figure}
%\centering
%\includegraphics[scale=0.65]{ADI}
%\caption{Bayes factor comparing the isothermal and the TP profile retrievals of the thermal inversion simulations (orange) and the non-thermal inversion simulations (blue). Bayes factor for TP profile are 10 to 70 times larger compared to isothermal model suggesting very strong preference for the TP profile models.}
%\label{fig: bayes-factor}
%\end{figure}

\subsubsection{With thermal inversion}
\label{inversion}

\subsubsection*{Thermal inversion without optical absorbers in the retrieval}

We now analyse with \taurex the spectra from the \textit{Thermal Inversion} simulations but ignoring the existence of TiO and VO as absorbers. TiO and VO were included in the GCM simulations but are not included in the forward radiative transfer model of \pytmo and their abundance is thus not a parameter to be retrieved, contrarily to CO and \hho{}.

\begin{figure}
\centering
\includegraphics[scale=0.6]{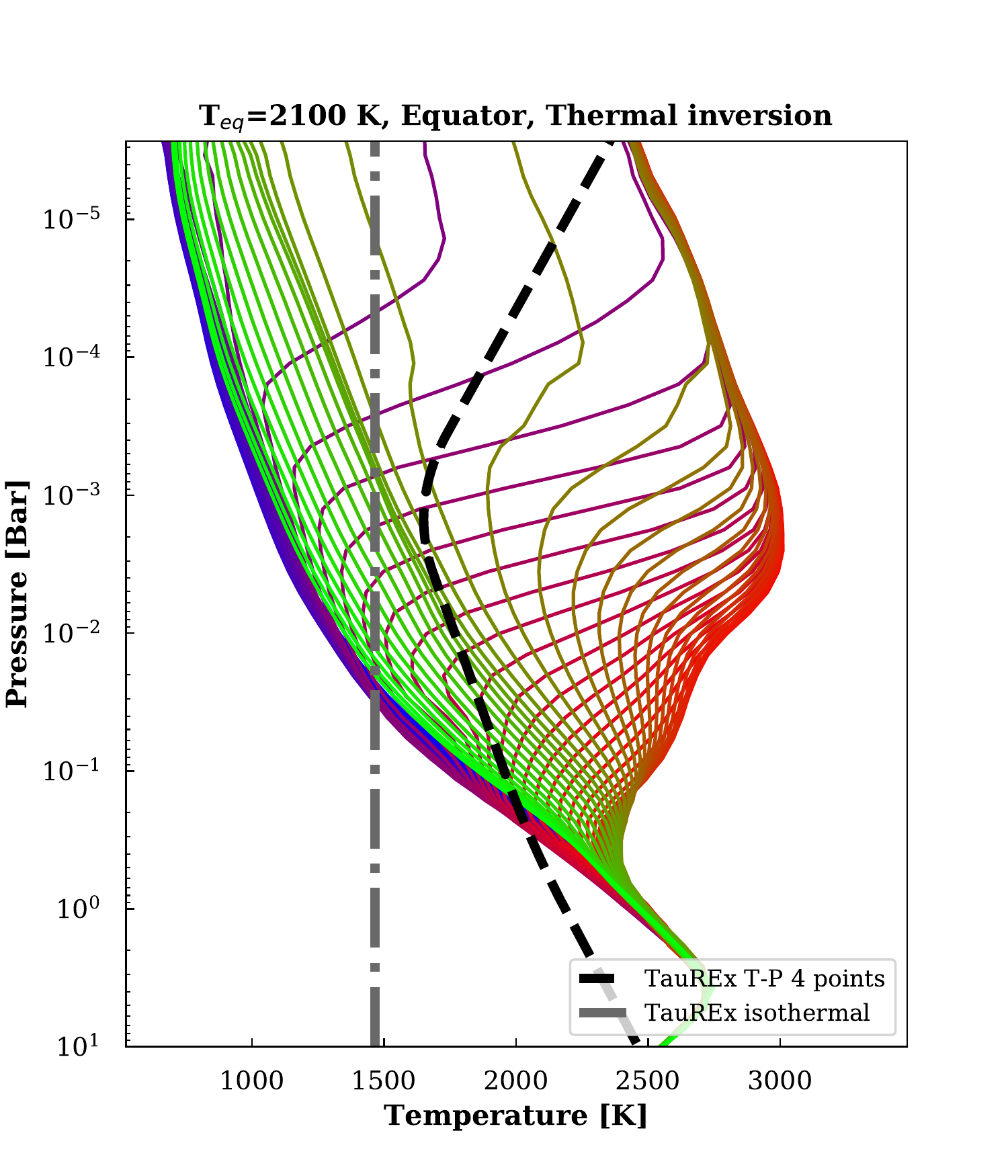}
\includegraphics[scale=0.35,trim = 0cm 0cm 7cm 0cm,clip]{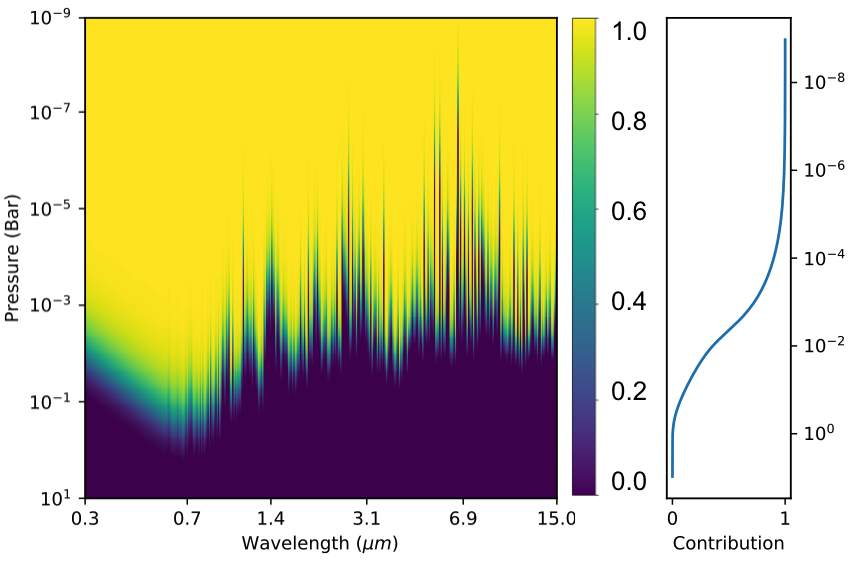}
\caption{\textbf{Top:} Equatorial thermal profiles from the \textit{Thermal Inversion} $T_\mathrm{eq}$\,=\,2100\,K simulation. Each color represents a longitude from anti-stellar (blue) to sub-stellar (red). The dotted-line grey (isothermal) and dashed-line black are the retrieved profiles. \textbf{Bottom:} Transmittance map (colorbar from 0 to 1) for each wavelength at each pressure.}
\label{fig: contrib-TiO}
\end{figure}

Unlike the previous section, the biases observed with the retrieved isothermal profiles remain with a 4-point thermal profile. Fig~\ref{fig: contrib-TiO} helps to understand why shaping the vertical profile does not improve the results. It shows the contribution function of the transmission spectra in the $T_\mathrm{eq} = 2100$~K simulation as well as the simulated and retrieved thermal profiles. We only plotted the GCM profile around the limb to focus on the probed region. Contrarily to the simulations without thermal inversion (Fig.~\ref{fig: contrib-noTiO}), the transition between the day and the night sides of the planet is sharper at the pressure below quasi-isothermal layer, the global gradient in order of magnitude (because that depends on the latitude, longitude and altitude) between the day and the night side is about 600~K/10$^\circ$ when it was more around 100~K/10$^\circ$ in the other simulations.
For this reason, it is impossible for the retrieval to find a 1D thermal profile which both fits with this simulation and find consistent values of the parameters.

The retrieved water abundances (Fig.~\ref{fig: comp-plot-4pts}) are now solar for every simulation and the CO abundances deviates more than the ones in the isothermal retrieval. As a result, the retrieved \COratio is slightly larger than the one inferred with the isothermal model. 
We also notice here a break in the slope at $T_\mathrm{eq} = 1700$~K: above this equilibrium temperature the biases increase more slowly than below.

Looking at the temperature and \hho{} abundance maps in Fig.~\ref{fig: maps-TiO}, we understand that for the coldest simulation ($T_\mathrm{eq} = 1400$~K) the spectrum is only sensitive to the shape of the thermal profile,
%dominated by the vertical evolution of the temperature, 
since the dissociation of \hho{} and \hh\ remains weak, hence a retrieved solar \COratio. Then, from $T_\mathrm{eq} = 1500$~K to $T_\mathrm{eq} = 1700$~K, the east-west asymmetry of the atmosphere combined with increasing thermal dissociation in the day side yields a large increase in the overestimation of the retrieved abundances by a combination of horizontal effects both along and through the limb. Finally, above $T_\mathrm{eq} = 1700$~K, although the thermal dissociation of the H$_2$O and H$_2$ intensifies, the east-west asymmetry becomes almost non-existent, which could explain the break in slope observed in the retrieved \COratio. For these hottest simulations, only horizontal effects through the limb seem to dominate the transmission spectra. Indeed, even if the retrieval using a non-isothermal profile fits the spectra better, the observed biases remain. 

The retrieved thermal profiles are shown in Fig.~\ref{fig: T-P_4pts}. \taurex favors a thermal inversion in all cases.

We performed the same statistical analysis as in Sec.~\ref{isothermal} with these simulations with \textit{Thermal inversion} but ignoring the existence of TiO and VO as absorbers. We calculated the logarithmic Bayes factor (Eq.~\ref{eq:logbayes_factor}). We present the value for each simulation in Table~\ref{tab: bayes-factor}. They indicate, as in Sec.~\ref{isothermal} with this simulations with \textit{No thermal inversion} that \taurex strongly favors the model with a 4-point TP profile compared to the isothermal model. The 4-point profile is even more suitable as the equilibrium temperature increases. While $\redchid$ was always greater than 1 in the isothermal retrieval, it is now below 1 for each simulation. Therefore, if the fits could have been rejected in this first case, \taurex tells us here that the fits are much better.

\subsubsection*{Thermal inversion with optical absorbers in the retrieval}

Finally, we studied what happens in more complex atmospheres where TiO and VO are present in addition to CO and \hho{}. Retrieval results, presented in Fig.~\ref{fig: comp-plot-4pts} (blue), remain close to those with an atmosphere containing only CO and \hho{}.

\COratio is still biased, all the more with hotter atmospheres. Differences appear in the retrieved abundances which are globally not solar in every simulation. TiO and VO abundances are under-estimated in the coldest simulations and they are close to the solar abundances in the hottest. This behavior can be explained by the condensation of these species that decreases their observed VMRs.

Plus, absorption features of VO in low resolution are hidden by the TiO bands, hence they weakly constrain the retrieval that reaches the limits of the priors (Fig.~\ref{fig: priors-TiO}). CO abundances are still overestimated, but less so than in the previous retrievals (one order of magnitude less in average). However, retrieved \hho{} abundances are underestimated, which explains the biased \COratio.

It is interesting to note that allowing the code to add TiO and VO which actually are present and affect the spectra does not lead to a better agreement, except partially for CO. The \hho{} abundances retrieved are even worse. We suppose it to be due to the model having a smaller margin to find a degenerated 1D solution. We also note that even when using a 4-point TP profile, $\redchid$ is above 1 for $T_\mathrm{eq}$ in the 1500--2100~K range. Thus a $\redchid$ test in these cases could be used as a warning.

As only CO does not dissociate or condense, we also plotted \VOratio and \TiOratio to see how far they are from the expected solar abundances. They are both biased, especially in the coldest simulations but that is due to both the condensation and the fact that VO features are hidden by TiO features in low resolution.

We conclude that 1D retrieval models are able to retrieve abundances within 1-$\sigma$ of their actual values in warm atmospheres, where the atmospheres are homogeneous in latitude and longitude. The errors bars estimated by \taurex\ underestimate however the departure. In the studied atmospheres, the transmission spectra are dominated by vertical effects at the limb which can be well reproduced by 1D models. However, these models cannot correctly reproduce the complexity of the 3D structure of hot exo-atmospheres, starting from $T_\mathrm{eq} \approx 1500$~K. The biases highlighted by \citet{Pluriel2020} therefore cover a larger number of objects, from HJs to UHJs (see Sec.~\ref{discussion} for more details). Moreover, if metals such as Fe or Mg, and ionized hydrogen are present in the atmosphere (as discussed in Sec.~\ref{GCMs}), these species would increase the atmospheric dichotomy, hence the magnitude of 3D effects, thus pushing to even colder equilibrium temperatures the 1D-model safe zone.

We also conclude that the model with a non-isothermal vertical profile is not relevant for the study of \wasp\ with $T_\mathrm{eq} \approx 2350$~K \citep{Pluriel2020}, where the horizontal heterogeneities through the limb have a greater contribution on the transmission spectrum than those due to vertical differences. Even if the Bayes factor of the simulations had been better with a 4-point thermal profile, we have shown here that it would not modify the biases observed. It makes sense to do a simpler retrieval with fewer parameters since the posterior distributions are very similar.

\section{How 3D effects affect 1D retrievals?}
\label{discussion}

The three-dimensional structure of HJ and UHJ atmospheres strongly impacts transmission spectroscopy and will bias the 1D retrieval models used to analyze and interpret future observational data from JWST and Ariel. These spectra carry information coming from various regions of the atmosphere, and it is often difficult to disentangle them. The 3D structure implies variations in the physical and chemical properties of the atmosphere, which affect the transmission spectrum, along three main axes: (i) variations as a function of altitude, i.e., the vertical effects; (ii) the north/south/east/west variations which we refer to as horizontal effects along the limb; (iii) and the variations between the day side and the night side, also referred to as horizontal effects through the limb. It is by ranking the impacts of these three contributions that we can determine how biased the retrieval models are. In this work, we have quantitatively characterized the biases observed, thus allowing a more exhaustive understanding of the effects involved, and we have highlighted the origins, as well as the limits of these biases. We present in Fig.~\ref{fig: plot-pedagogoc} a summary of our study where we suggest to the community which types of retrieval should be used depending on the equilibrium temperature of the planet and the presence or absence of optical absorbers in the atmosphere. 

\subsection{Vertical effects}

We highlight here the impact of the physical and chemical variations with altitude. These are the effect currently considered in atmospheric studies since most of them assume that the probed area in transmission remains a thin annulus around the terminator \citep{Tinetti2007,Redfield2008,Tsiaras_2018,Skaf2020,pluriel-ares2020}. 1D retrieval models are able to reproduce vertical effects with good accuracy, since they are able to retrieve vertical TP profiles. Therefore, when the transmission spectrum is mainly affected by temperature and/or chemical variations with altitude, models such as \taurex manage to accurately fit the observations and derive physical and chemical characteristics consistent with the input simulation or observation.

We demonstrated this with the retrieval on simulations without thermal inversion (Sec.~\ref{no inversion}). As a reminder, in these cases, the planet atmospheres remain relatively homogeneous and do not present a very strong day-night dichotomy, as shown in Fig.~\ref{fig: T-maps-noTiO}. Thus, on the one hand, the regions probed by the transmission spectra remain close to the terminator, and on the other hand this terminator is fairly homogeneous with slight east-west variations. It is then the vertical effects at the terminator which dominate the spectra shapes, thus allowing relevant and accurate 1D retrievals (Fig.~\ref{fig: comp-plot-4pts}). However, we also demonstrated that the isothermal hypothesis on the retrieval model should no longer be used for planets having an equilibrium temperature larger than 1000~K as it results in the inference of wrong abundances despite a \textit{good fit} ($\redchid \leq 1$).

1D retrieval models therefore remain suitable for studying not too hot Jupiter atmospheres ($T_\mathrm{eq} \leq 1400$~K), where horizontal variations along and through the limb can be neglected. However, we must remain aware that the region probed in this case mainly comes from a thin layer around the terminator. To determine the characteristics of the entire atmosphere, we need to compare these results with GCM simulations which model the entire atmosphere.

Finally, we find that when the atmosphere presents a very hot stratosphere ($T_\mathrm{eq} \geq 1500$~K), what controls most the shape of spectral features are no longer vertical gradients but horizontal ones. The results of the retrieval departs from the actual values far beyond estimated error bars and despite an excellent spectral match ($\redchid \leq 1$). This indicates that 1D vertical models provide an unrealistic solution to the observed spectra with good fits but with biased \hho{}, VO, TiO and CO abundances.

\subsection{Horizontal effects along the limb}

A second geometric effect can affect the shape of transmission spectra: differences along the limbs, in particular between the east and west limbs. This occurs when strong jets are present in the atmosphere or when the atmosphere is in super-rotation. This then creates an east-west asymmetry with sometimes important temperature differences, which significantly affects the spectrum.
This effect is highlighted in Figs.~\ref{fig: maps-TiO} and \ref{fig: T-maps-noTiO}, which compare the west and east limbs of 2 simulations, a cold one ($T_\mathrm{eq} = 1400$~K) and a hot one ($T_\mathrm{eq} = 2100$~K). They show an east limb that is generally colder than the west limb. The temperature difference becomes significant for the simulation with a hot stratosphere (with TiO/VO in the atmosphere)  down to $T_\mathrm{eq} = 1700$~K.
%, since the simulations have an west--east asymmetry as shown in the maps in Fig. \ref{fig: maps-TiO}. 
This is a clue to explain the break in slope in the retrieved \COratio ratios, which is observed in Fig.~\ref{fig: comp-plot-4pts}.

Several teams have studied east-west heterogeneities, whether in simulated or real observations. In particular, \citet{Lineparmentier2016} showed that the presence of a non-uniform cloud layer at the level of the terminator can affect the interpretation of transit observations. \citet{Powell2019} studied the impact of an heterogeneous cloud cover on the limbs of HJs, showing that the difference of the cloud properties between the east and the west limbs impacts the transmission spectra. GCM simulations also highlight these east-west asymmetries at the terminator, which seem to be consistent with observations by \citet{Cooper2008}. \citet{MacDonald2020} also studied the biases generated by these east-west differences and demonstrated their importance on the results of retrieval models. To take into account this heterogeneity at the terminator also allowed them to explain the unexpected cold temperatures retrieved  for WASP-17\,b and WASP-12\,b exoplanets.

It would be possible to solve this problem by increasing the temporal resolution between the ingress and the egress of the transit in order to differentiate at least two spectra, one coming from the eastern limb, the other from the western limb. Thus, we would end up with only the vertical differences that 1D data inversion models can handle.
Another way to obtain separately the information on each limb would be to analyze the phase curve. If the phase curve contained the observations of the transit ingress and egress with sufficient sampling, we could obtain separate transmission spectra for the east and west limbs.

\subsection{Horizontal effects across the limb}

We focus here on the effect of the day-night thermal and chemical gradients on the transmission spectrum. These horizontal effects are very often overlooked assuming that the transmission method only probes a thin annulus at the terminator, which is verified only for small enough atmospheric scale heights (see Fig.~2 by \citet{caldas2019}).
%in since the opacity decreases exponentially with altitude. Plus, the curvature of the atmosphere on the star-observer axis implies that the light generally probes a thin annulus around the terminator.
As shown by \citet{Pluriel2020}, we can see the inflated day-side atmosphere of UHJs and how they affect the geometry of the observation in Fig.~\ref{fig: maps-TiO}: the regions probed during transit is extended on the day side and is neither thin nor centered on the terminator. In such planets, the vertical effects become negligible compared with horizontal gradients since the day-night temperature contrast reaches thousands of kelvins while vertical temperature variations probed by the spectrum do not exceed hundreds of kelvins. In addition, gradients along the limb remain weak for the warmest atmospheres, hence a dominance of the effects through the limb in this case.

With night-side signatures hidden by the inflated day-side and mixed signatures at different temperatures and compositions in the observed spectrum, this configuration causes major biases in the properties retrieved by 1D retrieval models. 
%which are not able to correctly take into account the horizontal geometric variations where the inflation of the atmosphere on the day side of these planets hides the night side. 
Thus, whatever the nature of the assumed vertical profiles  (isothermal or not), 1D retrieval models fail to reproduce atmospheres faithful to the input GCM models. Furthermore, the thermal dissociation of species adds complexity to these geometric considerations, implying that the transmission spectrum probes a broad range of regions depending on the wavelength.

The impact of the day-night dichotomy on the transmission spectrum remains negligible for cold enough atmospheres, since for most of them they are not hot enough to present detectable dichotomy. On the other hand, when we study warm and hot atmospheres, we need to consider it because this effect can become dominant in the observations and therefore lead to erroneous interpretations. We note that there is an atmosphere regime (around planet with $T_\mathrm{eq}$\,=\,1700\,K) where the three effects described above are of the same order of magnitude, making the analysis of their transmission spectra extremely complex.

\begin{figure*}
\centering
\includegraphics[scale=0.45]{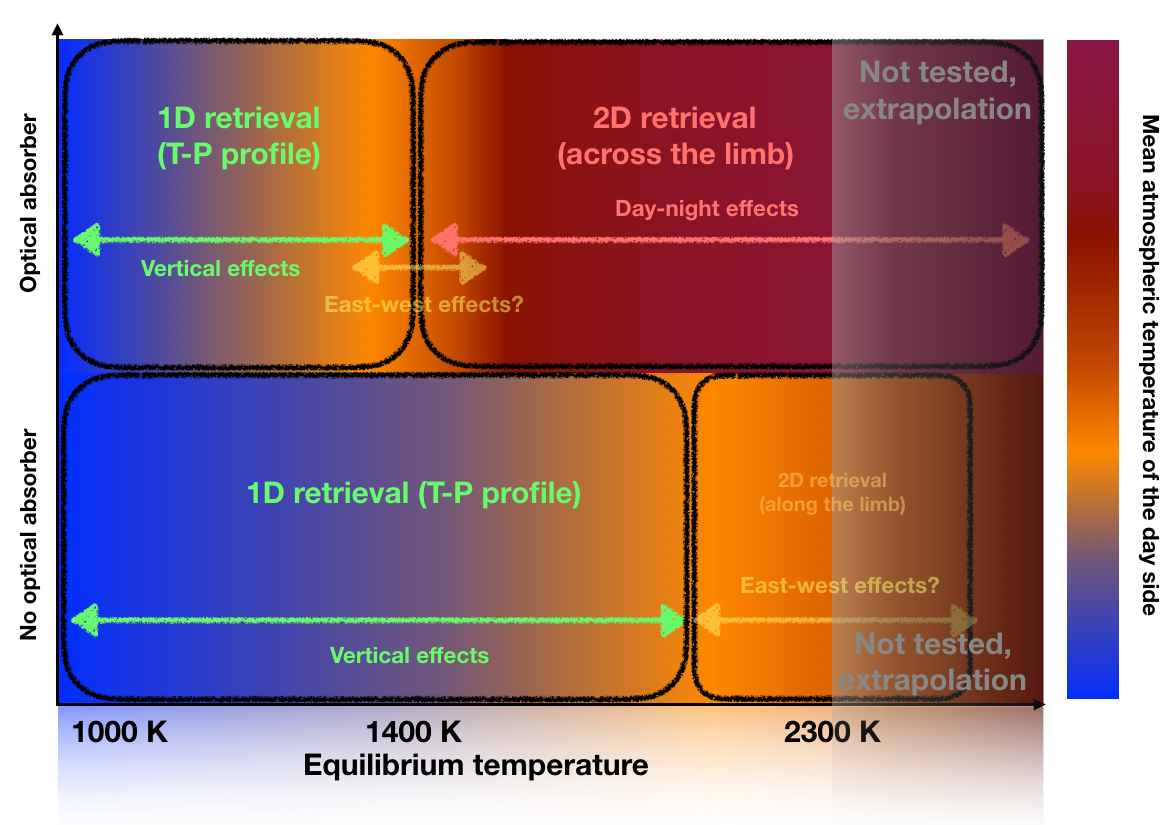}
\caption{Summary of the different geometries required in retrieval codes to avoid biases as a function of the equilibrium temperature of the planet and the presence/absence of optical absorbers (hence thermal inversion). 1D retrieval models appear to provide a relatively satisfactory parameter estimates for planets with equilibrium temperatures lower than 1400\,K when optical absorbers (TiO, VO, K, Na, metals, ionized hydrogen, etc) are present in the atmosphere. However, they lead to biased parameter estimates above this limit where 2D/3D retrieval codes are mandatory. When no optical absorbers are present in the atmosphere the validity of 1D retrieval code extends to an equilibrium temperature of 2000~K. Above this temperature, the estimated parameters become biased, probably due to east-west effects. Although we suggest clues about the impact of the east-west asymmetry, further investigations are needed to quantify their effects.}
\label{fig: plot-pedagogoc}
\end{figure*}

\section{Conclusions}

We have demonstrated that the shortcomings of 1D retrievals in the interpretation of transmission spectra are not limited to UHJs and can affect cooler planets as well. The 1D assumption in retrieval models will be an issue to accurately estimate the molecular abundances with future observations provided by JWST and Ariel. 

In particular, we have shown that the isothermal assumption leads to wrong retrieved abundances in every case we studied, even though the retrieved spectrum fits well the observational data. This means that isothermal atmospheres could give suitable spectrum fits but with very wrong abundances. Thus, we encourage the community not to use this assumption anymore when studying a planet with an equilibrium temperature larger than $T_\mathrm{eq}$\,=\,1000\,K. While parametrized thermal profiles yield retrieved abundances much closer to the actual abundances in the simulation, we nevertheless note that they produce inaccurate results for very hot atmospheres. This limit is reached for $T_\mathrm{eq}\ge 2100$~K, and even down to 1500~K when optical absorbers create a thermal inversion. 
Our findings confirm that these biases are mainly due to the strong day-to-night dichotomy, as shown by \citet{Pluriel2020}.

Based on our findings, \fig{fig: plot-pedagogoc} summarizes our recommendations on the minimal model assumptions necessary to avoid wrong interpretations and biased retrieved parameters. It can be used as follows:
\begin{enumerate}
\item Estimate/calculate the equilibrium temperature of the planet. 

\item Check from a first analysis of the data or estimate if the atmosphere is expected to contain optical absorbers (TiO, VO, K, Na, metals, ionized hydrogen...) which could create a thermal inversion.

\item Adapt the retrieval analysis and its interpretation according to \fig{fig: plot-pedagogoc}.
\end{enumerate}

If $T_\mathrm{eq}\geq 2100$~K, the parameter values (molecular abundances in particular) and their associated errors derived from a 1D retrieval are very likely to be wrong. A different retrieval method that accounts for the horizontal structure is then needed.
If $T_\mathrm{eq}\leq 1400$~K, the 1D hypothesis yields consistent parameter values and a 1D retrieval analysis can be used.
If $1500\:\mathrm{K} \leq T_\mathrm{eq} \leq 2100\:\mathrm{K}$, either (i) there are no optical aborbers in the atmosphere and in this case the 1D retrieval can lead to consistent parameter values, or (ii) there are optical absorbers in the atmosphere, so a hot stratosphere is likely present and the parameter values (molecular abundances in particular) inferred from a 1D retrieval procedure, as well as their estimated uncertainties, are very likely to be wrong. In case (ii), we suggest to use a different retrieval framework taking into account 2D effects across the limb.

We consider than 3D retrieval models based on GCM simulations would face two main issues: firstly, we would need massive computation power for Bayesian inference, secondly, we would have to deal with numerous degeneracies inherent to 3D structures. It would be extremely complex to break these degeneracies, even with the resolution and the accuracy of JWST or Ariel.

We therefore suggest to develop 2D retrieval models with a horizontal parametrization across the limb to be able to address the unavoidable imprint of horizontal gradients on the spectra. We think that 2D retrieval models have the right level of complexity for this task, therefore we developed a 2D retrieval code \citep{falco2021taurex}. We will study its impact on UHJs in a forthcoming paper.

\begin{acknowledgements}
We thank Michiel Min and Quentin Changeat for the useful discussions about the Bayesian statistical analysis.
This project has received funding from the European Research Council (ERC) under the European Union's Horizon 2020 research and innovation programme (grant agreement n$^\circ$679030/WHIPLASH). We thank the Programme National de Planétologie (CNRS/INSU/PNP) and the CNES for their financial support.
\end{acknowledgements}

\bibliographystyle{aa}
\bibliography{biblio}

\onecolumn
\appendix

\section{Appendix: Nested sampling posteriors}
\label{appendix:a}

We present here one Nested sampling posteriors for the T$_{\mathrm{eq}}$\,=\,2100\,K case with a thermal inversion in the GCM having H2O, CO, TiO and VO as absorber in the atmosphere. This is a typical corner plot from our study. As we did 28 retrievals, it would not have been relevant to put every Nested sampling posteriors in the paper, that's why we focused only on the mean value retrieved within 1-$\sigma$ error in Figs. \ref{fig: comp-plot-iso} and \ref{fig: comp-plot-4pts}.

\begin{figure}[h!]
\centering
\includegraphics[scale=.25]{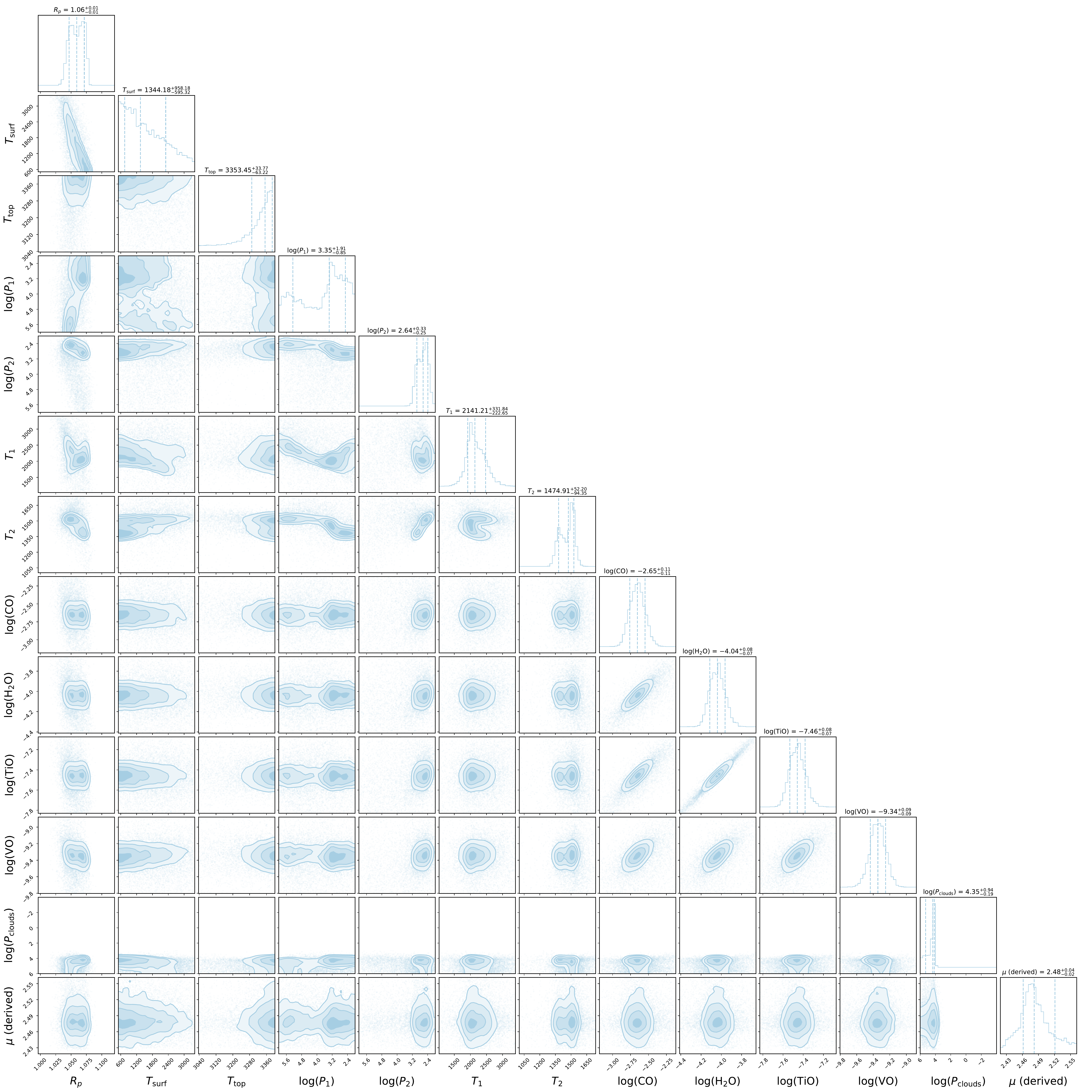}
\caption{Nested sampling posteriors retrieval from \taurex for the thermal inversion simulation considering TiO and VO as absorber in the atmosphere. We retrieved 12 free parameters, which are \hho{}, CO, TiO and VO abundances in log10(VMR), clouds pressure in Pa, 4 temperature points in Kelvin, 2 pressure points (in Pa) and the planetary radius in Jupiter's radius. The molecular weight is derived from these parameters. $T_{1}$ and $T_{2}$ are the temperature points corresponding the pressure points $P_{1}$ and $P_{2}$ respectively.}
\label{fig: priors-TiO}
\end{figure}

\end{document}